\documentclass[twocolumn,showpacs,preprintnumbers,
amsmath,amssymb,aps,prd,nofootinbib,superscriptaddress,
eqsecnum]{revtex4}
\usepackage{graphicx,color}

\makeatletter
\renewcommand{\p@subsection}{}
\makeatother
\renewcommand{\thefootnote}{\fnsymbol{footnote}}
\renewcommand{\thefootnote}{\#\arabic{footnote}}

\def\lsim{\mathrel{\rlap{\lower3pt\hbox{\hskip1pt$\sim$}}
     \raise1pt\hbox{$<$}}} 
\def\gsim{\mathrel{\rlap{\lower3pt\hbox{\hskip1pt$\sim$}}
     \raise1pt\hbox{$>$}}}
\def\be{\begin{eqnarray}}\def\ba{\begin{eqnarray}}
\def\ee{\end{eqnarray}}\def\ea{\end{eqnarray}}
\def\ben{\begin{enumerate}}\def\bitem{\begin{itemize}}
\def\een{\end{enumerate}}\def\eitem{\end{itemize}}
\def\Tr{{\rm Tr}}
\def\del{\partial}
\def\calL{\cal L}\def\calO{\cal O}

\def\pr{Phys. Rev.}\def\prl{Phys. Rev. Lett.}
\def\np{Nucl. Phys.}\def\pl{Phys. Lett.}

\def\bi{\bibitem}
\def\vA{{\mathbf{A}}}

\def\vJ{{\mathbf{J}}}

\def\vV{{\mathbf{V}}}
\def\vr{{\mathbf{r}}}
\def\vq{{\mathbf{q}}}
\def\vk{{\mathbf{k}}}
\def\vp{{\mathbf{p}}}
\def\vbp{{\overline {\mathbf{p}}}}
\def\vx{{\mathbf{x}}}
\def\vs{{\mathbf{\sigma}}}

\def\hatr{{\mathbf{{\hat r}}}}

\newcommand{\e}{{\mbox{e}}}
\def\fm{{\mbox{fm}}}

\def\nlo#1{\mbox{N$^{#1}$LO}}
\def\Sunit{\mbox{$10^{-20}$ keV-b}}
\def\calM{{\cal M}}

\def\voO{{\mathbf{{\cal O}}}}
\def\voT{{\mathbf{{\cal T}}}}

\def\dR{{\hat d^R}}
\def\fm{{\mbox{fm}}}
\def\vx{{\vec x}}
\def\EM{{\rm EM}}

\def\abs#1{{\left| #1 \right|}}

\def\nlo#1{{\mbox{N$^{#1}$LO}}}
\def\MS{{\mbox{M1V}}}
\def\mut{{\mbox{M1S}}}
\def\Qt{{\mbox{E2S}}}

\def\hatk{\hat{k}}
\def\pislash{ {\pi\hskip-0.6em /} }

\begin{document}


\title{
Predictiveness of Effective Field Theory in Nuclear Physics }

\author{Mannque Rho}
 \affiliation{ Service de Physique
Th\'eorique,  CEA Saclay, 91191 Gif-sur-Yvette c\'edex, France\\
(E-mail: mannque.rho@cea.fr)}

\date{\today}

\begin{abstract}
We discuss the role effective field theory plays in making
predictions in nuclear physics in an approach that combines both the
high sophistication of the standard nuclear many-body approach and
the power of systematic higher chiral-order account in chiral
perturbation theory. The main idea of this approach is illustrated
with a selected number of cases involving few-body systems, the
measurement of some of which poses an experimental challenge and
will be of value to solar neutrino studies.
\end{abstract}

\renewcommand{\thefootnote}{\#\arabic{footnote}}
\setcounter{footnote}{0}

\maketitle


\section{Introduction}
\label{sec:int}
Effective field theory is supposed to be an approach which can
ultimately reproduce a fundamental theory and hence in principle can
enable one to compute systematically corrections to approximate
calculations. In nuclear physics we are supposed to have a
fundamental theory, say, QCD, and so the task is to set up the
scheme which comes closest to the truth encoded in QCD. There is the
obvious question as to whether nuclear physics can indeed be
understood in terms of QCD or put differently, how to test QCD in
nuclear physics. By now this question has become akin to asking
whether condensed matter systems can be understood in terms of QED.
In this article, we adopt Weinberg's ``folk
theorem"~\cite{weinberg-folktheorem}~\footnote{Let us quote what
Weinberg says (the use of italic is the author's): ``When you use
quantum field theory to study low-energy phenomena, then according
to the folk theorem you're not really making any assumption that
could be wrong, unless of course Lorentz invariance or quantum
mechanics or cluster decomposition is wrong, provided you don't say
specifically what the Lagrangian is. As long as you let it be the
most general possible Lagrangian consistent with the symmetries of
the theory, you're simply writing down the most general theory you
could possibly write down.  ... Effective field theory was first
used in this way to calculate processes involving soft $\pi$ mesons,
that is, $\pi$ mesons with energy less than about $2\pi F_\pi\approx
1200$ MeV.  The use of effective quantum field theories has been
extended more recently to nuclear physics where although nucleons
are not soft they never get far from their mass shell, and for that
reason can be also treated by similar methods as the soft pions.
Nuclear physicists have adopted this point of view, and I gather
that they are happy about using this new language because it allows
one to show in a fairly convincing way that {\it what they've been
doing all along (using two-body potentials only, including one-pion
exchange and a hard core) is the correct first step in a consistent
approximation scheme}.... "} and eschew this issue entirely.

So what is the task of an effective field theory in nuclear physics?

In addressing this question, there are two possible philosophical
attitudes to take. One is to set up a well-defined effective theory
within a very restricted domain of applicability, say, in
energy/momentum or in the number of nucleons involved in the system
and investigate whether and how the intended theory does what it is
supposed to do. Here one has a well-defined set of rules for
calculations and then follow $rigorously$ the rules and confirm that
the strategy works. We shall call this ``rigorous EFT" (RigEFT in
short). In this approach, given the strong constraint imposed by the
consistency with the strict chiral counting and the rapid increase
of input parameters at higher orders, one is mostly limited to low
orders of chiral perturbation expansion, and trying to explain the
complex dynamics of wide-ranging nuclear systems is not presently
feasible. In the same class of approach is what we might call ``toy
EFT" where an EFT with the least complexity taken into account is
solved fully consistently with the set of rules adopted. The
``pionless EFT" ($\pislash$EFT) to be described below belongs to
this subclass. The other philosophy -- which in some sense is
drastically different from the $\pislash$EFT -- is, following
Weinberg's theorem, to take it for granted that EFT should work in
nuclear physics and exploit the power of EFT to make calculations
that cannot be accessed solely by the standard many-body technique
which has been developed since many decades. This is a more physical
approach aiming at describing nuclear phenomena more widely than
RigEFT. We shall call this ``more effective effective theory"
(MEEFT) to suggest that it exploits both the power of EFT and the
precision of the standard nuclear physics approach (SNPA in short).
\section{SNPA and EFT}
\subsection{The power of SNPA}
The standard approach to physics of finite nuclei and normal nuclear
matter has been to determine first highly sophisticated
phenomenological two-nucleon potentials implemented with
multi-nucleon (typically three-nucleon) potentials fit to a large
body of experimental data and solve the many-body problem as
accurately as possible. This approach, on a microscopic level, has
reached an impressively precise description of nuclei up to mass
number $A=10$ \cite{wiringaetal} with ground and excited state
energies within $\sim 1$ - 2\% and on a more macroscopical level, to
heavy nuclei. The solution requires massive numerical computations,
which some think render the approach inferior to or less elegant
than analytical approaches. But in the present era of computer
revolution, that is not a valid objection as it does not make the
work any less fundamental than analytical calculations. This
approach which we shall call ``standard nuclear physics approach
(SNPA)" exploits two-nucleon potentials that fit scattering data up
to momenta $\sim 300$ MeV with a $\chi^2$/datum $\lsim 1.4$. The
success of this approach has been extensively reviewed in
\cite{SNPA1,SNPA} and we will not elaborate any further on the SNPA
per se, although we will refer to it throughout this chapter. {\it
It should be stressed that it would be a grave mistake to ignore the
accuracy achieved by the SNPA and discard it as is done in some
circle of workers in the field on the ground that it is not derived
in EFT formalism. Our point of view which we will develop here is
that we should -- and can -- incorporate it in the framework of an
EFT in such as way as to allow us to systematically control possible
corrections brought in by more fundamental and systematic
formulation.}
\subsection{The power of EFT}
In assessing both the power and the limitation of the SNPA, the
physical observables to look at are not only the spectra of the
states involved but also response functions involving wave functions
and currents. For the latter, the standard procedure that has been
mostly employed is to write down the currents in terms of the
single-particle operators -- called impulse approximation -- which
are typically the most important and then make, often less
important, corrections due to the presence of multi-body operators
constructed from the exchange of mesons, called ``meson-exchange
currents," in an effective phenomenological Lagrangian field theory.
This approach was first systematized in \cite{CR71}. In many of the
applications, this approach gives results that compare well with
experiments. But within the framework of many-body theory based on
phenomenological potentials, however, there is no unique or
systematic way to assess what the size of corrections is, so when
the calculated value disagrees with experiments, there is no
well-defined and systematic way to improve the calculation. One is
thus allowed to make post-dictions but rarely {\it predictions} or
calculations that are free of free parameters. This is one place
where EFT can come in to help.

What is the true use of an EFT? To show that QCD works is not the
objective as mentioned above in accordance with Weinberg's folk
theorem. In essence QCD should work in nuclei if one works hard
enough. This we are witnessing in the construction of two-body
potentials in $\chi$PT.~\footnote{To be more specific, we note from
an updated review~\cite{machleidt06} that the $N^3$LO chiral
perturbation calculation of the two-body potential fits the 1999
data base $np$ scattering below 290 MeV with a $\chi^2$/datum=1.10
while the sophisticated phenomenological potentia AV18 fits it with
$\chi^2$/datum=1.04. To see how high-order $\chi$PT terms work, one
can compare the $\chi^2$/datum=36.2 and 10.1 respectively for NLO
and NNLO. At present, the $N^3$LO calculation is the best one can do
and it is highly unlikely that the SNPA potential be superseded by
high-order chiral perturbation.}
What have been done in the literature in the past in addressing the
issue in question are the following:
 \bitem
 \item One direction belonging to the RigEFT class
followed by a large number of workers in the field is to limit
oneself to a well-defined, but drastically simplified, EFT
Lagrangian
and then study this as rigorously as possible within the framework
of suitably well-defined rules. A case that has been widely studied
is the so-called ``pionless ($\pislash$)" Lagrangian approach in
which $all$ but nucleon fields are integrated out (e.g.,
\cite{pionless-chenetal} and for a review, see
\cite{bedaque-vankolck}). Covering by fiat a limited range of
nuclear interactions, it is necessarily limited in scope and so far
achieved no more than $reproducing$ what is already well and
accurately described in SNPA. While this approach has the advantage
to directly address problems that can lead to exact statements, such
as the phenomenon of renormalization-group limit cycle, Efimov
effect etc. (see, e.g., \cite{hammer}), it lacks predictiveness that
one would hope for.

An approach that belongs to the RigEFT class but comes closer in
principle to the MEEFT class takes the nucleon and the pion as the
relevant degrees of freedom and calculate to as high an order as
feasible and as consistently as possible within the tenet of chiral
perturbation theory. This calculation is limited in that the number
of unknown constants increases rapidly as the chiral order
increases. We will not have much to say on this approach in this
article. But when we say RigEFT without specification, we will mean
both this and the pionless EFT.
 \item
The other direction which we shall adopt in this paper as explained
in detail elsewhere~\cite{MR-taiwan} is to exploit the strategy of
an EFT to calculate quantities that neither the SNPA nor QCD proper
{\it separately} can do, that is, to make ``predictions." By
predictions, here, we mean parameter-free calculations with error
estimates of what might be left out in the approximations, e.g., the
truncation, involved. The strategy is to exploit both the full
machinery of SNPA $and$ the power of EFT in a scheme that is
consistent with the symmetries of QCD in the spirit of Weinberg's
theorem to make $predictions$ that can be confronted with Nature. We
will tacitly assume that systematic high-order chiral perturbation
calculations of the potentials will eventually provide quantitative
support to the potentials used in the SNPA. Up to date, this
assumption is justified as summarized in ~\cite{machleidt06}.
 \eitem
\section{Chiral Lagrangians}
\subsection{Relevant Scales and Degrees of Freedom}
When applied to nuclear systems, an EFT involves a hierarchy of
scales in the interactions. In nuclear physics, the nucleon is the
core degree of freedom defining the system with the pion figuring as
the lightest mesonic degree of freedom. What other degrees of
freedom must enter in the dynamics depends upon what problem one is
looking at. If the kinematics probed is of the scale $E\ll m_V$
where $m_V$ is the mass of the light-quark vector mesons, the
lightest of which are the $\rho$ and the $\omega$, then one may
``integrate out" all the vector mesons and work with the baryon and
the pion only. This regime is accessible by a chiral perturbation
theory ($\chi$PT) for physics of dilute nuclear systems, e.g.,
few-nucleon systems. If one works to high enough order in chiral
perturbation theory, one can hope to understand much of nuclear
physics taking place within the given restricted kinematic domain,
although in some cases, it is more convenient and simpler to
introduce heavier degrees of freedom explicitly as for instance the
$\rho$, $\omega$, $a_1$, $\sigma$ etc. When one is studying systems
involving a scale $E\ll m_\pi$, where $m_\pi$ is the pion mass, then
one might even ``integrate out" the pion as well, as is done in
$\pislash$EFT. How this program works is described in numerous
lecture notes, one nice reference of which is
\cite{kaplan-berkeley}.

It should be remarked that even when $\pislash$ EFT is fully
justified, that is, pions are $not$ indispensable, in certain cases,
it proves to be much more powerful and predictive to keep pions as
effective degrees of freedom. We will encounter such a situation
when we discuss the so-called ``chiral filter mechanism." One finds
that having explicit pion degrees of freedom with their associated
low-energy theorems can give a highly simplified and efficient
description of a process which would require much harder work if the
pion were integrated out.

More significantly, the explicit presence of pions is now known to
be the dominant element in describing the structure of light nuclei.
In particular, to quote Wiringa~\cite{wiringa06}, ``the success of
the simple model [of light nuclei] supports the idea that the
one-pion exchange is the dominant force controlling the structure of
light nuclei ..."\footnote{To continue the quote: ``In Green's
function Monte Carlo calculations of $A\leq 12$ nuclei with
realistic interactions, the expectation value of the
one-pion-exchange potential is typically 70-75\% of the total
potential energy. The importance of pion-exchange forces is even
greater when on considers that much of the intermediate range
attraction in the $NN$ interaction can be attributed to uncorrelated
two-pion exchange with the excitation of intermediate $\Delta
(1220)$ resonances. In addition, two-pion exchange between three
nucleons is the leading term in $3N$ interactions, which are
required to get the empirical binding in light nuclei. In
particular, the $3N$ forces provide the extra binding required to
stabilize the Borromean nuclei $^{6,8}$He and $^9$Be." It would seem
therefore that integrating out the pion is like first throwing away
the baby with the bathtub and then trying to recover the baby piece
by piece.}

As one probes denser systems such as nuclear matter and denser
matter, due to BR scaling (recently reviewed \cite{BHLR06}), the
vector meson mass drops and near chiral restoration, becomes
comparable to the pion mass. In this case, one cannot integrate out
the vector mesons; the vector fields have to be endowed with local
gauge invariance, so that systematic chiral perturbation can be done
with the vector mesons included \cite{HY:PR}.

\subsection{Vector mesons and baryons}
It has been recently discussed as to
how hidden local symmetry emerges if one wants to ``obtain" an
effective field theory from a fundamental theory.
%
Interestingly, a hidden local symmetry (HLS) involving an infinite
tower of gauge fields coupled to pions in a chirally symmetric way
emerges from string theory via holographic duality. One refers to
such a theory ``holographic dual QCD." Now one can view the hidden
local symmetry approach of Harada and Yamawaki~\cite{HY:PR} with the
light-quark vector mesons as a truncated version of the holographic
dual QCD. Restricted to the lowest members of the vector mesons,
i.e., the $\rho$ and $\omega$, it is the Wilsonian matching to QCD
that makes the theory a bona-fide effective theory of QCD. In the
simplest form of HLS theory with $N_f=3$, there are the octet
pseudo-Goldstone bosons and nonet of vector mesons coupled gauge
invariantly. Baryon fields do not figure explicitly. However
holographic dual QCD suggests that baryons must appear as skyrmions.
Although there have been efforts to construct skyrmions with a
chiral Lagrangian with vector mesons incorporated (see
\cite{weigel-IJMP} for references) -- and it is now clear that
vector mesons must be present in the skyrmion structure, very little
is understood of such hidden local symmetry skyrmions. For the
purpose of this section, instead of generating baryons as solitons,
we will simply introduce baryon fields explicitly.

\subsection{Baryon fields}
Baryon fields are to be treated as ``matter" fields, which is fine
as long as the momentum transfer involved is $\ll \Lambda$ where
$\Lambda$ is the cutoff. We generally consider baryons made up of
$u$, $d$ and $s$ quarks, i.e., $SU(3)_F$ although we will be
concerned only with two flavors in this article~\footnote{The
strange quark is found to figure less importantly than thought
before in the structure of the nucleon but for dense matter which is
one of the ultimate goals in nuclear physics, it should be
included.}. We want the baryon field $B$ to be invariant under
(vector flavor) $SU(3)_V$,
 \be
B\rightarrow VBV^\dagger
 \ee
with $V\in SU(3)_{L+R=V}$. Now how does $B$ transform under
$SU(3)_L\times SU(3)_R$? Here there is no unique way as long as it
is consistent with the symmetries of QCD. For our purpose, the most
convenient choice is to have it transform as
 \be
B\rightarrow hBh^\dagger
 \ee
with
 \be
\xi_L&\rightarrow& h\xi_L L^\dagger,\\
\xi_R &\rightarrow& h\xi_R R^\dagger
 \ee
where
 \be
U=\xi_L^\dagger \xi_R.
 \ee
Here $h$ is a complicated local function of $L\in SU(3)_L$, $R\in
SU(3)_R$ and $U\in SU(3)_V$, the explicit form of which is not
needed. When $L=R=V$, it is just a constant $h=V$.

We define in unitary gauge $\xi_L^\dagger=\xi_R=e^{i\pi/F_\pi}$
 \be
A_\mu &=& \frac i2 \left(\xi^\dagger \del_\mu\xi-\xi
\del_\mu\xi^\dagger\right),\\
V_\mu &=& \frac 12\left(\xi^\dagger \del_\mu\xi +\xi
\del_\mu\xi^\dagger\right), \\
D_\mu B &=& \del_\mu +[V_\mu, B]
 \ee
transforming under $SU(3)\times SU(3)$ as
 \be
A_\mu &\rightarrow& hA_\mu h^\dagger,\\
V_\mu &\rightarrow& h(\del_\mu + V_\mu) h^\dagger.
 \ee
Now what we need to do is to write the Lagrangian ${\cal L}_{inv}$
invariant and the Lagrangian ${\cal L}_{non}$ non-invariant under
the given transformation. Thus
 \be
{\calL}_{inv}&=& \Tr \bar{B}(i\gamma^\mu D_\mu - m_0)B -D\Tr
\bar{B}\gamma^\mu\gamma_5\{A_\mu,B\}\nonumber\\
 && -F\Tr\bar{B}\gamma^\mu\gamma_5[A_\mu,B] +\cdots, \\
{\calL}_{non} &=& a_1 \Tr\bar{B}(\xi^\dagger M\xi^\dagger + {\rm
h.c})B +a_2\Tr\bar{B}B(\xi^\dagger M\xi^\dagger+ {\rm
h.c.})\nonumber\\
&& +a_3\Tr(MU+{\rm h.c.})\Tr\bar{B}B +\cdots
 \ee
Here the ellipses stand for higher order terms, either in
derivatives or in quark mass terms or in combination of both. $m_0$
is the dynamically generated mass of the baryon which is of order
$\sim 1$ GeV, i.e., the chiral scale.

In nuclear physics at low energy we are considering, the nucleon is
non-relativistic, so it makes sense to go to the non-relativistic
form of the Lagrangian. In fact for the development of the principal
thesis of this article, it is preferable. A nice way of going to the
non-relativistic form is the heavy-baryon
formalism~\cite{heavybaryon}. To do this, define
 \be
B_v (x)=e^{im_0 v\cdot\gamma v\cdot x}B(x)
 \ee
and introduce the spin operator $S_v^\mu$ that satisfies
 \be
v_\mu S_v^\mu=0,\ S_v^2 B_v=-\frac 34 B_\nu,
 \ee
 and
 \be
\{S_v^\mu,S_V^\mu\}=\frac 12(v^\mu v^\nu-\eta^{\mu\nu}),
 \ee
 \be
[S_v^\mu,S_V^\nu]= i\epsilon^{\mu\nu\alpha\beta} v_\alpha
S_{v\beta}.
 \ee
In terms of these definitions, we can rewrite ${\calL}_{inv}$ as
 \be
{\calL}_{inv}= &&\Tr \bar{B}_vv_\mu D^\mu B_v +2D\Tr \bar{B}_v
S_v^\mu \{A_\mu,B_v\} \nonumber\\
&&+2F\Tr \bar{B}_vS_v^\mu [A_\mu,B_v] +\cdots
 \ee
In this form the mass term $m_0$ disappears, so the chiral counting
comes out as we wanted without having to worry about the cancelation
between the mass term and the time derivative on the baryon field.
Similar forms can be written down for the chiral symmetry
non-invariant terms.

The Lagrangians written above in bilinears in the baryon field are
applicable in the one-nucleon sector. With these Lagrangians in
various approximations, one can systematically treat one baryon
problem including interactions with pions and external fields. If
one is interested in this topic, there are some good reviews in the
literature which show that chiral perturbation theory does work well
and could even be improved as one goes to higher orders and as more
precise experimental data become available. In what follows and in
other sections of this paper, we will take this ``success" for
granted.

Here we are interested in few- as well as many-nucleon systems. When
there are more than one nucleon in the system, one has multi-nucleon
terms involving $2n$ Fermion fields for $n>2$. Thus we have
 \be
{\calL}_{2n-fermi}=\sum_a(\bar{B}\Gamma_{a\mu}
B)(\bar{B}\Gamma_a^{\prime\mu} B)+\cdots
 \ee
where the ellipses now stand for higher number of Fermi fields and
$\Gamma$ and $\Gamma^\prime$ are various Lorentz structures
involving derivatives etc. subject to the necessary symmetry
constraints of QCD.

If one wishes, the light-quark vector mesons can be incorporated in
a hidden-gauge symmetric way although there are certain ambiguities
which are not present in the skyrmion approach mentioned above. In
this article, we won't need vector degrees of freedom since their
masses are of higher scale than the appropriate cutoff in the
density regime we are considering.
\section{Pionless EFT}
As mentioned, if one is interested in nuclear processes where the
energy scale is much less than the pion mass, one may integrate out
the pion as well. The resulting Lagrangian containing only the
massive nucleons has no chiral symmetry since there are no pions
anymore. But chiral symmetry is not violated.  It is just that
chiral symmetry is an irrelevant symmetry here. It is easy to write
down the effective Lagrangian
 \be
{\calL}_{\not\pi} &=& N^\dagger (i\del_t +\nabla^2/2m)N \nonumber\\
&& + C(N^\dagger N)^2 +C^\prime ((NN)^\dagger (N\nabla^2 N)
+\cdots\label{pionless}
 \ee
where $N$ stands for the nucleon doublet replacing the baryon field
$B$. The ellipsis stands for terms with spin and flavor matrices and
other derivative terms of the same order with and without spin and
isospin factors etc. The Lagrangian should be Galilean invariant.

Given the extreme simplicity of the pionless Lagrangian, the power
counting rule is equally simple allowing one to do a systematic and
consistent calculation in principle to high orders.  It is just a
power series in $p/\Lambda$ where $p$ is the probe momentum and
$\Lambda$ is the cutoff scale defining the momentum space
considered. In practice it makes no sense to go to very high orders
since unknown parameters increase rapidly. Further complications can
arise if one is interested in many-nucleon systems where the Fermi
momentum which is of order of a few times the pion mass enters but
this theory makes no sense there anyway.

Now what can we learn with this? As already mentioned, actually
little more than that an EFT works for effective range theory. It
can work for two-body and perhaps three-body systems for which the
SNPA has scored a great success already but the procedure gets
rapidly unmanageable when it goes to the really interesting problem
like the solar $hep$ and the related $hen$ process that involve
four-body interactions. (We will return to this problem below.) It
has no pions and hence quantities which are rendered easy to
understand when pions are present (various low-energy theorems ..)
are made difficult, if not impossible (as we will see later). Being
a toy model, it allows systemization and completeness in describing
certain but restricted low-energy processes thanks to the simplicity
of the Lagrangian. But it should be mentioned that the physical
processes that can be treated are mostly, if not all, those which
have been well reproduced since ages in SNPA with well-defined
corrections (e.g., those processes which are subsumed in effective
range theory).

As an illustration, consider $n+p\rightarrow d+\gamma$ at low energy
which has been heralded as a success of the model
\cite{kaplan-berkeley} and to which we will return below for MEEFT.
The $np$ capture at thermal energy is accurately measured, and so
can offer an excellent process to check the theory with. Now this
process is dominated by an isovector $M1$ operator, so in
(\ref{pionless}), we need the interaction $C$ term for the $^3S_1$
channel, a $C^\prime$ term for both $^3S_1$ and $^1S_0$ channels and
the couplings to the magnetic photon
 \be
{\calL}^{1b}_{B}=eN^\dagger (\kappa_S
+\kappa_V\tau_3)\frac{{\mathbf\sigma}\cdot{\mathbf B}}{2m} N
 \ee
and
 \be
{\calL}^{2b}_{B}=eL_1 (N^T P_i N)^\dagger (N^T {\overline{P}_3}
N)B_i
 \ee
where $\kappa_{S/V}= \frac 12(\kappa_p \pm \kappa_n)$ with
$\kappa_p=2.79$ and $\kappa_n=-1.91$. Here $P_i$ and
$\overline{P}_3$ are, respectively, the $^3S_1$ and $^1S_0$
projection operators. The constant $C$ in (\ref{pionless}) is
obtained by fitting the deuteron binding energy and the $C^\prime$
from the effective range in $NN$ scattering. But there is one
unknown constant $L_1$ which cannot be a priori fixed: There is no
other experiment than the $np$ capture that involves this term and
hence no true prediction can be made. It is not however totally
devoid of value since once $L_1$ is fixed from the $np$ capture
experiment, one can then turn the process around and calculate the
inverse process $\gamma d\rightarrow n+p$ as a function of the
photon energy. It works fairly well up to, say, $E_\gamma \sim 10$
MeV. This is relevant for big-bang nucleosynthesis
~\cite{kaplan-berkeley}. One should however recognize that there is
no real gain in theoretical understanding in this approach because
the SNPA can do just as well with well-defined two-body corrections.
\section{MEEFT}
\subsection{Weinberg's counting rule}
We now turn to a more predictive EFT scheme. To exploit the accuracy
and power of SNPA, we adopt the Weinberg counting rule.\footnote{As
far as the author is aware, the first effort to incorporate Weinberg
counting in nuclear physics was made in 1981 \cite{MR-Erice81} based
on Weinberg's 1979 paper on $\pi\pi$
scattering\cite{weinberg-physica} which contains the core idea
behind the application of chiral perturbation theory to nuclear
problems. Unfortunately the discussion in \cite{MR-Erice81} was
incomplete because the role of nucleons in the counting was
inadvertently left out.} In fact the Weinberg
counting\cite{weinberg-counting1,weinberg-counting2} is the only way
that the SNPA can be ``married" with an EFT. Since the aim here is
to exploit the powers of both SNPA and EFT, we refer to this as
``MEEFT (more effective effective field theory)."

The Weinberg counting consists essentially of two steps. In the
first step, one defines the nuclear potential as the sum of
2-particle irreducible diagrams -- irreducible in the sense that
they contain no purely nucleonic intermediate states -- and
truncates the sum at some order $n$ in the standard chiral
perturbation power counting. The potential is dominated by two-body
irreducible terms with $n$-body terms with $n>2$ terms suppressed by
the power counting. In the second step, the potential so constructed
is used to solve the Lippman-Schwinger or Schr\"odinger equations.
This corresponds to incorporating ``reducible" graphs that account
for infrared enhancement (or singularities) associated with bound
states etc. Now the potential will then include one or more pion
exchanges and local multi-fermion interactions that represent
massive degrees of freedom (such as the vector mesons $\rho$,
$\omega$ etc.) that are ``integrated out." The approach can then be
applied not only to few-nucleon processes but also to many-nucleon
processes with the outputs in the form of physical observables,
namely, scattering cross sections, energy spectra, response
functions to external fields etc. Wave functions make an
indispensable part of the scheme, so the theory enables one to
calculate things not only in few-body systems but also across the
periodic table. Contact with ``rigorous" EFT -- in the sense of
power counting assured, e.g., in pionless EFT -- can be made only
for few-body systems. However systematic higher order chiral
corrections to the estimates made in SNPA can be made with certain
error estimates for processes that involve heavy nuclei, e.g., lead
nuclei.
This feat has not been feasible up to date in the $\pislash$EFT
approach

The Weinberg counting rule explains automatically that n-body forces
with $n > 2$ are suppressed relative to 2-body forces and n-body
currents are normally subdominant to one-body current, known in
nuclear physics as impulse approximation etc. These are familiar
things in SNPA but in MEEFT, they go beyond SNPA in that one can in
practice make a systematic account for corrections to the SNPA
results that are guided by an effective field theory strategy. What
this means is that in calculating response functions to external
fields, one can use the accurate state-of-the art wave functions
that have been constructed since a long time in SNPA to make
``predictions" that are otherwise not feasible in the RigEFT
approach. We will illustrate this point below.

There is of course a price to pay for this ``predictive power." One
of them is possible ambiguity in the regularization procedure which
separates high-energy and low-energy physics. Where to do the
separation is completely arbitrary and the observables calculated
should not depend on the separation point. This is the statement of
the renormalization group invariance. However to the extent that one
calculates in certain approximation, one cannot make the
independence perfect. As it stands at present, how best to do this
can be more an art than science.
\begin{figure}[th]
\centerline{\includegraphics[width = 7.5cm]{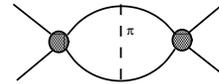}}\vskip
-3.5cm \caption{\label{2-loop-reg} A two-loop diagram that requires
higher-order counter term in the leading order graph in
$\pislash$EFT.}
\end{figure}

To cite a specific case, one of the (counting) problems arises when
the two-loop graph with a pion exchange between two nucleons of Fig.
\ref{2-loop-reg} treated as a Feynman diagram is calculated in
dimensional regularization. The problem lies in treating this graph
strictly in the sense of perturbation theory. We will see later that
this is $not$ what is done in MEEFT. If one writes the dimension as
$d=4-2\epsilon$ where we want $\epsilon\rightarrow 0$, then one
finds that this graph diverges as
 \be
\sim \frac{1}{\epsilon} m_\pi^2.
 \ee
This represents the logarithmic dependence on scale $\sim \ln\mu$.
This scale dependence needs to be canceled by a counter term that
goes like $\sim m_\pi^2$. In the power counting of the theory, this
comes at higher order than that assigned to the pion exchange in
Weinberg counting since $m_\pi^2$ counts as ${\calO} (Q^2)$. This
means that when one solves Schr\"odinger equation with the pion
exchange included in the potential consisting of irreducible terms,
one has an inconsistency in chiral counting since the pion exchange
needs to be regulated by a counter term higher order in the Weinberg
counting.

Now how did this ``basic" problem not hamper development in SNPA?
The answer is that what happens in nuclei is not wholly accessed by
perturbative renormalization that is obstructed by the above
problem. The fact that one is solving Schr\"odinger equation
illustrates that nuclear problems are inherently non-perturbative
and the above inconsistency arises because one is doing a
perturbation calculation. This difficulty is essentially sidestepped
by MEEFT that we will discuss in the following section.

There is also a possibility that this regularization difficulty can
even be $formally$ remedied~\cite{vankolcketal1,arriola1}. One of
the most serious is the singularity of tensor interactions in
partial waves where the interaction is attractive. For large cutoff
$\Lambda$, say, greater than vector meson mass, spurious bound
states can be generated. Of course the cutoff has a physical meaning
in effective theories as elaborated further below and picking a
cutoff bigger than what is appropriate is meaningless, so this
problem is in a sense academic. However even for low enough cutoff,
there may be unacceptably large cut-off dependence. The situation
gets more serious in certain channels where counter terms are
infrared-enhanced. However it turns out that one can add a counter
term in each partial wave where this sensitivity is present and
avoid this difficulty, at least to the leading
order\cite{vankolcketal1}. It is not clear that this problem
continues to arise at higher orders and more work is needed in this
area. This issue is revisited in \cite{epelbaum-meissner06}. At
present, the MEEFT predictions described below are free from this
difficulty.
\subsection{Strategy of MEEFT}
What is crucial for a viable MEEFT is to have very accurate wave
functions. Let us suppose that we do have wave functions that give a
good description of spectra and response functions for a range of
mass numbers. {\it The basic requirement is that the wave functions
so obtained possess the correct long-distance properties governed by
chiral symmetry.} We shall assume that the wave functions we use
meet this requirement. It would be highly surprising if fitting a
large number of low-energy data, both scattering and current matrix
elements, can be achieved without a good account of the
long-distance physics associated with chiral symmetry. We can
therefore assume that the currents we construct in Weinberg's scheme
incorporating one or more pion exchanges given by leading order
chiral expansion should be consistent with the (SNPA) wave functions
up to that order, with uncertainty in the current residing in higher
order. Shorter-distance properties of the range $ \gsim
(2m_\pi)^{-1}$ in the phenomenological potentials used to generate
the wave functions are not unique, but they should not affect long
wavelength probes we are interested in. If we can assure cutoff
independence within a range consistent with the physics we are
looking at, that should be good enough for a genuine prediction. One
could use renormalization group arguments to support this statement.
Now the strategy in computing response functions is to calculate the
irreducible contributions to the current vertex functions to the
order that is correctly implemented in the wave functions from the
point of view of chiral counting. In doing this, we can roughly
separate into two classes of processes: those processes that are
chiral-filter protected and those that are not. The former processes
are accurately calculable given accurate wave functions with small
error bars since the effects are primarily controlled by soft-pion
theorems. The latter is less well controlled by low-energy theorems
and hence brings in certain uncertainties. We will show however that
with an astute implementation of SNPA, one can calculate certain
processes belonging to the second class with some accuracy. We will
treat both classes below.
\subsection{The ``chiral filter"}
We will be dealing with a slowly varying weak external field, with
the external field acting only once at most. This means that in the
chiral counting, the external field that does not bring in small
parameters such as the external momentum or the pion mass lowers by
one chiral order two-body corrections relative to the
single-particle process to the counting that goes into the
potential. This is essentially because of the minimal coupling of
the external field which replaces one derivative in the two-body
terms by the external field that contribute to the potential. This
means that if we know the potential to $n$-th chiral order, then
there can be a two-body current which, {\it when not suppressed by
symmetry}, is determined to the same order, further corrections
being suppressed by two chiral orders. This makes the calculation of
the correction terms extremely accurate. This does not take place if
the external field brings into the current operators extra small
parameters such as the external momentum or the pion mass. It is
known that this chiral filter is operative with the isovector $M1$
and the axial-charge operators in nuclei\cite{KDR,MR91}, but not
with the isoscalar $M1$ and the Gamow-Teller operators as we will
see later.

The idea of chiral filter explained more precisely below was a
guiding principle when EFT formalism was not available for nuclear
physics. In chiral perturbation theory, chiral filter is automatic,
so there appears to be no big deal in it when one does a systematic
chiral expansion to high enough order but the intuitive power
associated with pion dynamics is completely lost when the pion is
integrated out as in the pionless EFT. An illustrative example is
the thermal $np$ capture which can be predicted without any
parameters in MEEFT but can only be postdicted with one unknown
parameter in the pionless theory. The point is that even if it were
possible to do a systematic chiral counting in $\pislash$EFT and
obtain certain results perhaps painlessly, the simplicity brought in
by chiral symmetry would be lost.
\subsection{Working of MEEFT}
In order to illustrate how MEEFT supplemented by the chiral filter
mechanism works out in nature, let us study the response functions
in the mass number $A$ system where we have at our disposal accurate
wave functions for $A-1$, $A$ and $A+1$ systems. We shall do so with
few-nucleon systems. For heavy nuclei and nuclear matter, we cannot
proceed in the same approach and hope to obtain accuracy. For that
purpose, we need to develop the notion of ``double decimation" as we
shall mention in the last section.

We want to treat specifically $A=2, 3, 4$ systems on the same
footing. The processes we are interested in are
 \be
n+p&\rightarrow& d+\gamma,\nonumber \\
p+p&\rightarrow& d+e^+ +{\nu}_e,\nonumber\\
\nu_e+d&\rightarrow& e^{-} + p+ p,\nonumber\\
e+d&\rightarrow& e+d,\nonumber\\
p+^3{\rm He}&\rightarrow& ^4{\rm He}+e^+ +\nu_e,\nonumber\\
n+^3{\rm He}&\rightarrow& ^4{\rm He} +\gamma.\label{meeft}
 \ee
These are processes that are of interest not only for nuclear
physics but also for astrophysical studies. The first four processes
were postdicted in the $\pislash$EFT with undetermined
parameters\cite{chen-np-supp,ravndal-pp} and the last two cannot be
accessed even for postdiction. The third process is one of the class
of reactions for SNO processes studied in
EFT~\cite{andoetal-neutrino}, involving both the charge current (CC)
processes
 \be
\nu_e+d &\rightarrow& e^{-} + p+ p,\nonumber\\
 \bar{\nu}_e +d &\rightarrow& e^+ +n+n\label{CC}
  \ee and the neutral current (NC)
processes
 \be
{\nu}_l +d &\rightarrow& {\nu}_l +p+n,\nonumber\\
\bar{\nu}_l +d &\rightarrow& \bar{\nu}_l+p+n.\label{NC}
 \ee
Here $l=\e, \mu, \tau$. These neutrino processes are very closely
related to the $pp$ process as they are governed by the same
operators, albeit at different kinematics. They can be $predicted$
with an accuracy comparable to that of the $pp$
process~\cite{andoetal-neutrino}. The fourth process,
electron-deuteron scattering, can be treated in various versions of
EFT with some accuracy for momentum transfer $Q^2 < 1$ GeV$^2$. All
the processes in (\ref{meeft}) have been treated in SNPA but for
reasons not difficult to pin-point, various calculations gave
different results ranging in some cases over several orders of
magnitude. The problem there was that there was no systematic way of
estimating errors involved and making corrections. We will see how
the chiral filter works in enabling one to make an accurate
(parameter-free) prediction for the isovector transition in the $np$
capture and how it provides a way to compute the processes that are
not protected by the chiral filter. The last of (\ref{meeft})-- the
$hen$ process -- has not yet been fully calculated in the approach
described in this paper. However the current is completely
determined with the formalism described here and there is nothing to
prevent a totally parameter-free calculation as we shall
argue.\footnote{A preliminary calculation performed in an ``MEEFT"
formalism (to be developed below) by Park and
Song~\cite{park-song-hen} suffers from a technical defect in the
description of the initial scattering state, unrelated to the main
issue of this paper, so the final numerical result cannot be
trusted. However the discussion of the EFT strategy is entirely
correct. The numerical result duly corrected will be forthcoming.}
\subsubsection{What can the chiral filter say?}
Historically it was the special role played by the pion that led to
an early and simple understanding~\cite{riska-brown} of the thermal
$np$ capture process in (\ref{meeft}). Now we know that the chiral
filter mechanism is automatically included in a systematic chiral
expansion in MEEFT, so there is no big deal in it from the point of
view of chiral symmetry but the power of it is in highlighting where
and in which way pions can play a prominent role in certain
processes, without requiring to go to higher orders in chiral
expansion and in avoiding unknown parameters. It has the power to
presage what one can or cannot do with calculable leading
corrections. This was first noticed in the application of current
algebras, i.e., soft-pion theorems, to the nuclear processes of the
type given in (\ref{meeft})\cite{KDR} and subsequently justified in
$\chi$PT~\cite{MR91}. The argument went as follows.

\begin{figure}[htb]
\vskip -0.0cm \center{\includegraphics[width = 9.5cm]{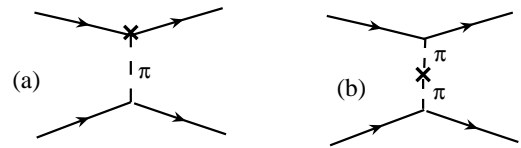}}
\vskip -3.5cm \caption{\small Two-body currents with one soft-pion
exchange which  dominate whenever unsuppressed by kinematics or
symmetry. The cross stands for the current. Both (a) and (b)
contribute to the vector current but only (a) contributes to the
axial current. }\label{onepi}
\end{figure}

In MEEFT, effective currents are characterized by the number of
nucleons involved in the transition. The leading one (in the chiral
counting) is the single-particle opeerator (called one-body or
impulse), then the next subleading one is two-particle (two-body)
followed by three-particle etc. The Weinberg counting says that in
the processes we are concerned with, we can limit ourselves to up to
two-body. It has indeed been verified that three-body and
higher-body corrections can be safely ignored\footnote{Contrast this
to the $\pislash$EFT where $n$-body currents for all $n>1$ appear at
the leading order.}. Now among two-body currents, the leading
correction will be given by the one-pion exchange as depicted in
Fig.\ref{onepi}.
This is the longest-range correction. Shorter-ranged corrections
involve higher derivative terms, pion loop diagrams and
corresponding counter terms. If the one-pion exchange contribution
can contribute unsuppressed by kinematics or symmetries, then we
have a chance to estimate the leading corrections with confidence
free of parameters. Since the $\pi NN$ vertex is known, what matters
then will be the vertex $J_\mu\pi NN$ where $J_\mu=V_\mu^a,\
A_\mu^a$ with the index $a$ standing for the isospin (flavor). Now
the longest wavelength process that enters here will be the graph
with the pion being ``soft." When the pion is soft, there is a
low-energy theorem that gives for the vector current
 \be
 &\sim& {\calO}(\frac{p}{m_N})\ \ \ \ \ \ {\rm for}\ \ \ \mu=0,\\
 &\sim&
 \frac{g_A}{F_\pi}\epsilon^{3ab}\frac{\tau^a}{2}{\mathbf{\sigma}} \ \ {\rm for}
\ \ \ \mu=1, 2, 3
  \ee
and for the axial vector current
 \be
&\sim& \frac{1}{F_\pi}\epsilon^{abc} V_0^c\ \
{\rm for}\ \ \ \mu=0,\\
&\sim& {\calO} (\frac{p}{m_N}) \ \ \ \ \ \ {\rm for}\ \ \ \mu=1, 2,
3.
 \ee
Here $m_N\sim 1$ GeV is the lowest baryon, i.e., nucleon mass. The
characteristic momentum scale involved, say, momentum carried by the
nucleon responding to the external field, is assumed to be much
smaller than the baryon mass scale. What these results imply is
quite simple. They say that the two-body corrections are dominated
by the soft-pion exchanges in the space component of the vector
current, e.g., the isovector $M1$ transition and in the time
component of the axial vector current, e.g., the first forbidden
beta transitions $J^\pm\rightarrow J^\mp$ with $\Delta T=1$. As a
corollary, we learn that the two-body currents for the time
component of the vector current, e.g., the charge operator and the
space component of the axial current, e.g., the Gamow-Teller
transition have no reason to be unsuppressed. In contrast, the
leading order one-body current has the opposite behavior, namely,
the space component of the one-body vector current and the time
component of the one-body axial current are suppressed relative to
the other components. Table \ref{table-counting} illustrates what
sorts of chiral orders are involved for the vector current in this
way of counting. A similar scaling applies to the axial current.

\begin{table}
\begin{center}
{\begin{tabular}{ccc}
 \hline
     & $V^i$ & $V^0$ \\
\hline
one-body current & ${\calO} (Q^{1})$ & ${\calO} (Q^{0})$\\
two-body (leading)current & ${\calO} (Q^{2})$& ${\calO} (Q^{3})$ \\
two-body (one loop)current & ${\calO} (Q^{4})$& ${\calO} (Q^{4})$ \\
$M_{2b}$/$M_{1b}$&${\calO} (Q^{1})$& ${\calO} (Q^{3})$\\
\hline
\end{tabular}}
\end{center}\caption{Various contributions to the vector current relative to
the one-body charge $V^0\sim {\cal O}(1)$. }\label{table-counting}
\end{table}

\subsubsection{Sketch of the calculational procedure}
While the calculation involves a conceptually simple procedure, the
details are rather complicated. We shall therefore skip the details
and instead present a brief sketch of the calculational procedure
focusing more on the essential concepts.

In the Weinberg counting scheme, the relevant quantity is the index
$\nu$ in the chiral counting of the electroweak currents. In the
present case it is sufficient to focus on ``irreducible graphs" in
Weinberg's classification. Irreducible graphs are organized
according to the chiral index $\nu$ given by
 \be \nu = 2 (A-C) + 2 L
+\sum_i \nu_i\,, \label{nu}
 \ee
where $A$ is the number of nucleons involved in the process, $C$ the
number of disconnected parts, and $L$ the number of loops; $\nu_i$
is the chiral index $\nu \equiv d+e+\frac{n}{2}-2$ of the $i$-th
vertex where $d_i$ is the number of derivatives, $e$ the number of
the external field ($=1$) and $n_i$ the number of internal nucleon
lines, all entering the $i$-th vertex. One can show that a diagram
characterized by eq.(\ref{nu}) involves an $n_B$-body transition
operator, where $n_B \equiv A- C+1$. The physical amplitude is
expanded with respect to $\nu$. The leading-order one-body GT
operator belongs to $\nu$=0. Compared with this operator, a Feynman
diagram with a chiral index $\nu$ is suppressed by a factor of
$(\breve{q}/\Lambda_\chi)^\nu$, where $\breve{q}$ is a typical
three-momentum scale or the pion mass, and $\Lambda_\chi \sim$ 1 GeV
is the chiral scale. We denote this in short as $Q^\nu$.
In our case it is important to take into account also the kinematic
suppression of the time component of the nucleon four-momentum. We
note
 \be
v\cdot p_l \sim v\cdot p_l' \sim v\cdot k_l \sim
\frac{\breve{q}^2}{m_N}, \label{kinetic}
 \ee
where $p_l^\mu$ ($p_l'^\mu$) denotes the initial (final) momentum of
the $l$-th nucleon, and $k_l^\mu \equiv (p_l'-p_l)^\mu$. Therefore,
each appearance of $v\cdot p_l$, $v\cdot p_l'$ or $v\cdot k_l $
carries two powers of $\breve{q}$ instead of one, which implies that
$\nu$ increases by two units rather than one. Thus, if we denote by
$q^\mu=(q_0,\mathbf{q})$ the momentum transferred to the leptonic
pair, say,  in the $pp$ and $hep$ processes in eqs.(\ref{meeft}),
then $q_0\sim |\mathbf{q}|$$\sim q^2/\Lambda_\chi$ $\sim{\cal
O}(Q^2)$ rather than ${\cal O}(Q)$ as naive counting would suggest.
These features turn out to simplify the calculation considerably.

In what follows, unless otherwise stated, we will always count
correction terms relative to the leading-order terms. Since the
leading-order terms in the currents are of ${\cal O}(Q^0)$, a chiral
order corresponding to the index $\nu$ will often be referred to as
\nlo{\nu}; $\nu$=1 corresponds to NLO (next-to-leading order),
$\nu$=2 to N$^2$LO (next-to-next-to-leading order), and so
on.\footnote{In our notation, the power $\nu$ in N$^\nu$LO
represents the factor $Q^\nu$ relative to the leading order terms
$\sim {\cal O}(1)$ which should not be confused with other
conventions found in the literature. These two notations will be
used interchangeably in this paper.} In this discussion, we shall
limit ourselves to \nlo3. One can go to \nlo4 for certain operators
as was done in the published calculation~\cite{hep} but this is not
needed for our purpose here.

We now briefly sketch the structure of one-body (1B) and two-body
(2B) current operators that are consistent with the chiral counting.
Of course actual calculations have to go into the nitty-gritty
details.

The current in momentum space is written as
 \be J^\mu(\vq)= V^\mu(\vq) + A^\mu(\vq)
 = \int\! d\vx\, \e^{- i\vq\cdot \vx} J^\mu(\vx).
\ee  We shall use the obvious notations $ V^\mu=(V^0,\,
\vV)\,,\;\;\;\; A^\mu=(A^0,\,\vA)$.

The chiral counting of the electroweak currents is summarized in
Table \ref{power}, where the non-vanishing contributions at $\vq=0$
are indicated. This table shows how the counting goes for each
component of the currents:
\begin{itemize}
\item
$\nu$ = 0 : One-body $\vA$ and $V^0$: $\vA$ gives the Gamow-Teller
(GT) operator, while $V^0$ is responsible for the charge operator.
\item
$\nu$ = 1 : One-body $A^0$ and $\vV$: $A^0$ gives the axial-charge
operator while $\vV$ gives the M1 operator.
\item
$\nu$ = 2 : Two-body tree current with $\nu_i=0$ vertices, namely,
the soft-pion-exchange current. This is the leading correction
protected by chiral filter to the one-body M1 and axial-charge
operators carrying an odd orbital angular momentum.
\item
$\nu$ = 3 : Two-body tree currents with $\sum_i \nu_i=1$. These are
leading corrections to the GT and $V^0$ operators carrying an even
orbital angular momentum. These are chiral-filter $unprotected$ and
hence involve constants that are not given by chiral symmetry
considerations (e.g., soft-pion theorems) alone.
\item
$\nu$ = 4 : All the components of the electroweak current receive
contributions of this order. They consist of two-body one-loop
corrections as well as leading-order (tree) three-body corrections.
Among the three-body currents, however, there are no six-fermion
contact terms proportional to $(\bar N N)^3$, because there is no
derivative at the vertex and hence no external field.
\end{itemize}
\begin{table}[bth]
 {\begin{tabular}{|c||l|l|l|l|l|}
\hline $J^\mu$ & LO & NLO & \nlo2 & \nlo3 & \nlo4
\\ \hline
$\vA$ & 1B & $-$ & 1B-RC & 2B & 1B-RC, 2B-1L and 3B \\
$A^0$ & $-$ & 1B & 2B & 1B-RC & 1B-RC, 2B-1L \\
$\vV$ & $-$ & 1B & 2B & 1B-RC & 1B-RC, 2B-1L \\
$V^0$ & 1B & $-$ & $-$  & 2B & 1B-RC, 2B-1L and 3B \\ \hline
\end{tabular}}\caption{\protect Contributions from each type of current at $\vq=0$.
The entry of ``$-$" indicates the absence of contribution. ``1B-RC''
stands for relativistic corrections to the one-body operators, and
``2B-1L'' for one-loop 2-body contributions including counter term
contributions. }\label{power}
\end{table}

In this subsection we will deal with the isovector currents,
returning to the isoscalar currents later.

One can easily see that the counting rule for $\vV$ is the same as
for $A^0$, and similarly for $V^0$ and $\vA$. Table~\ref{power}
illustrates in what way $\vV$ and $A^0$ are chiral-filter-protected
while $V^0$ and $\vA$ are not. The essential feature is encapsulated
in the ratio $2B/1B\sim {\cal O} (Q^1)$ dictated by chiral symmetry
for the former vs. $2B/1B\sim {\cal Q} (O^3)$ for the latter for
which chiral symmetry has little, if any, to say.

We now show the explicit expressions for the relevant currents. This
is to give an idea what sorts of operators are involved in view of
the general discussion given above.  Up to \nlo3, the 1B currents in
coordinate representation are well-known in the literature,
 \be \tilde V^0(l) &=& \tau_l^- \e^{-i \vq\cdot \vr_l}
\left[
 1 + i \vq \cdot \vs_l \times \vp_l
\frac{2 \mu_V -1}{4 m_N^2}
 \right],
\nonumber \\
\tilde \vV(l) &=& \tau_l^-  \e^{-i \vq\cdot \vr_l} \left[
 \frac{\vbp_l}{m_N}\left(1 - \frac{\vbp_l^2}{2 m_N^2}\right)
 + i \frac{\mu_V}{2 m_N} \vq \times \vs_l
 \right.\nonumber\\
 &&\left.
 + i \vs_l\times \vbp_l\, q_0 \frac{2 \mu_V-1}{4 m_N^2}
 \right],
\nonumber \\
\tilde A^0(l) &=& - g_A \tau_l^- \e^{-i \vq\cdot \vr_l} \left[
 \frac{\vs_l\cdot \vbp_l}{m_N}
 \left(1 - \frac{\vbp_l^2}{2 m_N^2}\right)
 \right],
\nonumber \\
\tilde \vA(l) &=& - g_A \tau_l^- \e^{-i \vq\cdot \vr_l} \left[
 \vs_l
 \right.\nonumber\\
 &&\left.
 + \frac{2 (\vbp_l\, \vs_l \cdot \vbp_l - \vs_l \, \vbp_l^2)
     + i \vq\times \vbp_l}{4 m_N^2}
 \right],
 \label{J1Bnon}\ee
where $\mu_V \simeq 4.70$ is the isovector anomalous magnetic moment
of the nucleon and $\tau_l^- \equiv \frac12 (\tau_l^x - i\tau_l^y)$.
The tildes in eq.(\ref{J1Bnon}) imply that the currents are given in
the coordinate space representation, and $\vp_l = -i \nabla_l$ and
$\vbp_l = -\frac{i}{2} \left( \stackrel{\rightarrow}{\nabla}_l -
\stackrel{\leftarrow}{\nabla}_l\right)$ act on the wave functions.

We next consider the two-body (2B) currents. Because of the chiral
filter protection, the two-body operators $\vV_{\rm 2B}$ and
$A^0_{\rm 2B}$ are determined unambiguously. These have no unknown
parameters. This means that processes dominated by these operators
can be calculated parameter-free. The $V^0_{\rm 2B}$ operator does
not appear up to the order under consideration, so we will forget
it. The two-body currents that concern us are given in the
center-of-mass (c.m.) frame in coordinate space with the cutoff
incorporated as specified below by
 \be \vV_{12}(\vr)
&=& - \frac{g_A^2 m_\pi^2}{12 f_\pi^2} \tau_\times^- \,
 \vr \,
 \left[
 \vs_1\cdot\vs_2 \, y_{0\Lambda}^\pi(r)
 + S_{12} \, y_{2\Lambda}^\pi(r) \right]
\nonumber \\
 &-& i \frac{g_A^2}{8 f_\pi^2}
 \vq\times \left[
 \voO_\times y_{0\Lambda}^\pi(r)
 +\left( \voT_\times - \frac23 \voO_\times \right)
  y_{1\Lambda}^\pi(r)
 \right],
\nonumber \\
A^{0}_{12}(\vr) &=& - \frac{g_A}{4 f_\pi^2} \tau_\times^- \left[
\frac{\vs_+ \cdot \hatr}{r} + \frac{i}{2} \vq\cdot \hatr\,
\vs_-\cdot \hatr\, \right] y_{1\Lambda}^\pi(r) ,
\nonumber \\
\vA_{12}(\vr) &=& - \frac{g_A m_\pi^2}{2 m_N f_\pi^2} \Bigg[
\nonumber \\
&&\left[ \frac{\hat c_3}{3} (\voO_+ + \voO_-) +\frac23 \left(\hat
c_4 + \frac14\right)
   \voO_\times \right] y_{0\Lambda}^\pi(r)
\nonumber \\
&&
  + \left[
     \hat c_3 (\voT_+ + \voT_-)
  - \left(\hat c_4 + \frac14\right) \voT_\times
     \right] y_{2\Lambda}^\pi(r)
   \Bigg]
\nonumber \\
&+& \frac{g_A}{2 m_N f_\pi^2 } \Big[ \frac{1}{2} \tau_\times^-
   (\vbp_1 \,\vs_2\cdot\hatr +
   \vbp_2\,\vs_1\cdot\hatr)
\frac{y_{1\Lambda}^\pi(r)}{r}
\nonumber \\
&&
 + \delta_\Lambda(r)
 \, \hat d^R \voO_\times
\Big], \label{vAnuFT}
 \ee
where $\vr=\vr_1 - \vr_2$, $S_{12}= 3 \vs_1\cdot \hatr\,\vs_2\cdot
\hatr - \vs_1\cdot\vs_2$, and $\voO_{\odot}^{k} \equiv \tau_\odot^-
\sigma_\odot^k$, $ \voO_{\odot} \equiv \tau_\odot^- \vs_\odot$, $
\voT_{\odot} \equiv
 \hatr\, \hatr\cdot \voO_{\odot} - \frac13 \voO_{\odot}$,
$\odot=\pm,\times$, $\tau_\odot^-\equiv(\tau_1\odot\tau_2)^-\equiv
(\tau_1\odot\tau_2)^x -i (\tau_1\odot\tau_2)^y$ and
$\vs_\odot\equiv(\vs_1\odot\vs_2)$.
%
%
It is important to note that in (\ref{vAnuFT}), it is only in
$\vA_{12}(\vr)$ that undetermined parameters appear. We will show
how they can be fixed unambiguously from the accurate experiment on
the triton beta decay. This is the crucial point of this approach
that renders it truly predictable.

Note that the vector current is completely free of parameters.

\subsubsection{How the cutoff $\Lambda$ enters}
The two-body currents derived usually in momentum space are valid
only up to a certain cutoff $\Lambda$. This implies that, when we go
to coordinate space, the currents must be regulated. With the cutoff
having a physical meaning, there are no divergences and no
perturbative counter term problem discussed by the aficionados of
the $\pislash$EFT. This is a key point in our approach.  In
performing Fourier transformation to derive the $r$-space
representation of a transition operator, we can use a variety of
different regulators and physics should not be sensitive to the
specific form of the regulator. A simple and convenient
regularization is the Gaussian form
 \be S_\Lambda(\vk^2) =
\exp\!\left(\!-\frac{\vk^2}{2\Lambda^2}\!\right). \label{regulator}
 \ee
In terms of this function, the regularized delta and Yukawa
functions take the form
 \be
\delta_\Lambda^{(3)}(r)&\equiv&
 \int \!\!\frac{d^3 \vk}{(2\pi)^3}\,
  S_\Lambda^2(\vk^2)\, \e^{ i \vk\cdot \vr}\,,
 \nonumber\\
  y_{0\Lambda}^\pi(r)&\equiv&
  \int \!\!\frac{d^3 \vk}{(2\pi)^3}\,
  S_\Lambda^2(\vk^2)\, \e^{ i \vk\cdot \vr}
  \frac{1}{\vk^2 + m_\pi^2}
  \nonumber\\
y_{1\Lambda}^\pi(r) &\equiv& - r \frac{\del}{\del r}
y_{0\Lambda}^\pi(r)
  \nonumber\\
y_{2\Lambda}^\pi(r) &\equiv & \frac{1}{m_\pi^2}
  r \frac{\del}{\del r} \frac{1}{r}
  \frac{\del}{\del r} y_{0\Lambda}^\pi(r)\ .\label{reg}
 \ee

As noted above, the chiral-filter-protected operators
$\vV_{12}(\vr)$ and $A^{0}_{12}(\vr)$ are given by the soft-pion
exchange and hence will contain the Yukawa functions. Given the wave
functions, the matrix elements of these operators are unambiguously
given with marginal dependence on the cutoff due to the
shorter-ranged function $y_{1\Lambda}^\pi(r)$. On the other hand,
the chiral-filter-unprotected operator $\vA_{12}(\vr)$ contains, in
addition to the long-ranged Yukawa term $y_{0\Lambda}^\pi(r)$ and
the short-ranged $y_{2\Lambda}^\pi(r)$ with the fixed $\hat{c}$
coefficients, delta function terms containing the only parameter of
the theory $\dR$ as one can see from (\ref{vAnuFT}).

\subsubsection{Physical meaning of $\Lambda$} Unlike in a
renormalizable field theory where the cutoff is to be sent to
$\infty$, the cutoff parameter $\Lambda$ in EFT defines the physics
of the system we are interested in. In fact there is no strictly
renormalizable field theory known in the real world; the cutoff
always is finite in theories of the real world. In our case, a
reasonable range of $\Lambda$ may be inferred as follows. According
to the general {\it tenet} of $\chi$PT, $\Lambda$ larger than
$\Lambda_\chi \simeq 4\pi f_\pi \simeq m_N$ has no physical meaning.
In this sense, worrying about what happens when the cutoff is taken
to $\infty$ -- sometimes found in the literature of nuclear EFT --
is unwarranted~\cite{epelbaum-meissner06}. Meanwhile, since the pion
is an explicit degree of freedom in our scheme, $\Lambda$ should be
much larger than the pion mass to ascertain that genuine low-energy
contributions are properly included. These considerations lead us to
adopt as a natural range $\Lambda$ = 500-800 MeV, in the range where
the lowest-lying vector mesons intervene.

There is a subtlety in handling the delta-function appearing in the
two-body axial current (\ref{vAnuFT}) that we need to discuss before
proceeding further. This concerns specifically our later treatment
of the axial current, so let us consider the axial current. Later we
will see that a similar argument can be made for the chiral-filer
unprotected vector current. For definiteness, let us take $\Lambda$
= 500, 600 and 800 MeV. Now the procedure is that for each of these
values of $\Lambda$ one adjusts $\dR$ to reproduce the experimental
value of the triton decay rate $\Gamma_\beta^t$. To determine $\dR$
from $\Gamma_\beta^t$, one calculates $\Gamma_\beta^t$ from the
matrix elements of the current operators evaluated for accurate
$A$=3 nuclear wave functions. What is important is to maintain
consistency between the treatments of the $A$=2, 3 and 4 systems,
including the same regularization applied to all processes. In
numerical work in \cite{hep}, the Argonne $v_{18}$ (AV18)
potential~\cite{av18} for all these nuclei plus the Urbana-IX
(AV18/UIX) three-nucleon potential for the $A\ge 3$ systems were
used.
As long as the potential is ``realistic" in the sense that it has
the correct long-range part and fits scattering data well, it should
not matter what potential one uses. This is guaranteed by the RGE
argument on $V_{lowk}$ discussed in \cite{BR:DD,BHLR06}. If it were
to have an appreciable dependence, that would mean that the scheme
could not be trusted.

The values of $\dR$ determined in this manner are:
 \be \dR &=&
1.00\pm0.07\;\;\;\;\;\;\; {\rm for}\;\;\Lambda=500\; {\rm MeV}\,,
\nonumber\\
\dR &=& 1.78\pm0.08\;\;\;\;\;\;\; {\rm for}\;\;\Lambda=600\; {\rm
MeV}\,,
\label{dR}\\
\dR &= &3.90\pm0.10\;\;\;\;\;\;\; {\rm for}\;\;\Lambda=800\; {\rm
MeV}\,, \nonumber
 \ee
where the errors correspond to the experimental uncertainty in
$\Gamma_\beta^t$. These values determined in the three-body system
will be used below in both two-body and four-body systems.
\subsection{The MEEFT predicts}
\subsubsection{Thermal $np$ capture}
We first look at the case where a clean prediction can be made free
of short-distance ambiguity. It is the total cross section at
thermal energy for the process
 \be
n+p&\rightarrow& d+\gamma.\label{npcapture}
 \ee
Here the isovector M1 operator dominates. There are corrections from
the isoscalar currents (isoscalar M1 and E2, see
eq.(\ref{xsection})) but they are down by three orders of magnitude,
so to the accuracy involved, we can ignore them for the total cross
section. The calculation presented here represents a $\chi$PT
improvement on the earlier Riska-Brown work~\cite{riska-brown} based
on the Chemtob-Rho procedure~\cite{CR71}. In the next subsection, we
will discuss the polarization observables for which the isoscalar
currents figure prominently.

\begin{figure}[htb]
\centerline{\includegraphics[width = 7.5cm]{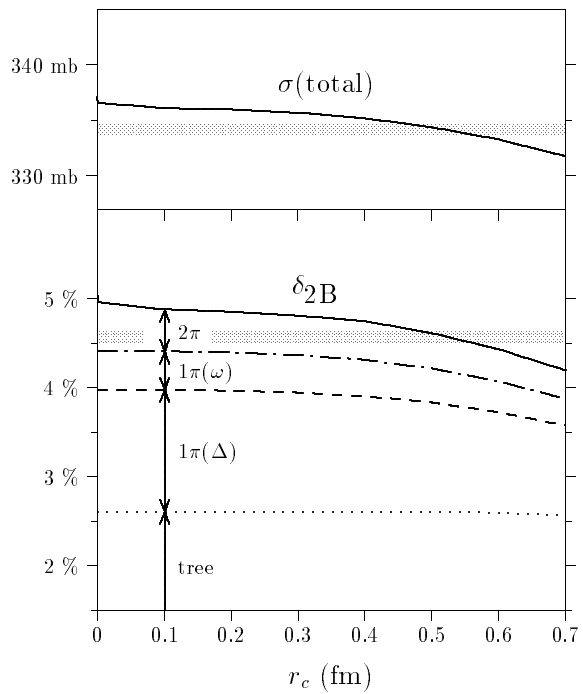}}
 \caption{\protect \small Total capture cross section $\sigma_{\rm
cap}$ (top) and the corrections relative to the single-particle M1
matrix elements denoted $\delta$'s (bottom) vs. the cut-off $r_c$.
The solid line represents the total contributions and the
experimental values are given by the shaded band indicating the
error bar. The dotted line gives $\delta_{\rm tree}$, the dashed
line $\delta_{\rm tree} + \delta_{1\pi}^\Delta$, the dot-dashed line
$\delta_{\rm tree} + \delta_{1\pi}= \delta_{\rm tree} +
\delta_{1\pi}^\Delta + \delta_{1\pi}^\omega$ and the solid line the
total ratio, $\delta_{\rm 2B}$.} \label{npresult}
\end{figure}

As first discussed in \cite{KDR},
the isovector M1 operator has the chiral filter protection. This
means that the principal correction to the leading one-body M1
operator of ${\calO}(Q)$ relative to the one-body term comes from a
one soft-pion exchange graph. One can verify that the next
correction comes at ${\calO}(Q^3)$. With the ${\calO}(Q^3)$ one-loop
corrections to the vertices, the chiral filter argument predicts
simply that the one-pion exchange with the parameters completely
fixed by chiral symmetry should dominate the two-body current. There
is a small ${\calO}(Q^3)$ correction coming from one-loop graph
involving two-pion exchanges. The result of the calculation which
involves only the cutoff to be taken care of, here represented by
the cut-off radius in coordinate space (referred in nuclear physics
to as ``hard-core radius"), is given in Fig.\ref{npresult} from
\cite{pmr-np1}. The various contributions correspond to the
following. The ``tree" corresponds to soft one-pion exchange term
(of ${\calO}(Q)$) with the constants at the vertices given by
renormalized quantities that can be picked from experiments. The
terms with $1\pi (\omega, \delta)$ represent one pion exchange of
${\calO}(Q^3)$ with the vertices resonance-saturated with the
$\omega$ and $\Delta$. The ``$2\pi$" represents 2$\pi$ exchange
one-loop corrections.

The remarkable point to note is the weak dependence on the cut-off
radius $r_c$, ranging from 0 to 0.7 fm. This is an a posteriori
check of the consistency of the procedure. Taking into account the
variation over this range as a measure of the error involved in the
calculation, the prediction is
 \be
\sigma_{th}=334\pm 2\ \ {\rm mb}\label{nptheory}
 \ee
to be compared with the experimental value
 \be
\sigma_{ex}=334.2 \pm 0.5.\ \ {\rm mb}\label{npexp}
 \ee
By going beyond the N$^2$LO (relative to the single-particle M1
matrix element), one could bring the accuracy of the theory within
the experimental uncertainty but this calculation has not been
performed yet.
 \subsubsection{Polarization observables in $np$
 capture}\label{isoscalar}
We now turn to the case where while the chiral filter protection is
not available, one can still make an accurate prediction. Together
with the $hep$ process discussed below, this represents a highly
non-trivial case where the MEEFT works surprisingly well. It
concerns the isoscalar matrix elements in the $np$ capture process
(\ref{npcapture}) which contribute insignificantly to the total
capture rate and hence are ignored in the theoretical value of the
cross section. The process involved is the polarized process
 \be
\vec{n}+\vec{p}\rightarrow d+\gamma.\label{polnp}
 \ee
We discuss how a parameter-free prediction on spin-dependent
observables of this process at threshold can be made. This
discussion is based on the work of \cite{pkmr-isonp}.

To bring out the main points of the calculation, we need to specify
in more detail than what we did above for the total capture rate
what sorts of matrix elements are involved. We shall be brief on
this, however.

The process (\ref{polnp}) receives, apart from the contributions
from the isovector M1 matrix element (M1V) between the initial
${}^1S_0$ ($T=1$) and the final deuteron ($T=0$) state, the
isoscalar M1 matrix element (\mut) and the isoscalar E2 (\Qt) matrix
element between the initial ${}^3S_1$ ($T=0$) and the final deuteron
${}^3S_1\!-\!{}^3D_1$ states. While the spin-averaged cross section
$\sigma_{unpol}(np\rightarrow d\gamma)$ is totally dominated by M1V,
since the initial ${}^1S_0$ state has $J=0$, the M1V cannot yield
spin-dependent effects, whereas \mut{} and \Qt{} can. This means
that the spin-dependent observables in (\ref{polnp}) are sensitive
to small isoscalar matrix elements.

Now recall that the isoscalar matrix elements, M1S and E2S, are
examples of the chiral-filter unprotected observable. Furthermore,
it is well known that the one-body contribution of \mut{} is highly
suppressed due to the orthogonality between the initial ${}^3S_1$
and the final deuteron state in addition to the small isoscalar
magnetic moment of the nucleon. The soft-pion exchange is also
suppressed, there being no isoscalar ${\cal B}^\mu\pi NN$ vertex in
the leading order chiral Lagrangian, where ${\cal B}^\mu$ is the
external isoscalar field that couples to the baryonic current. Due
to this double suppression, the size of $\mut$ becomes even
comparable to that of \Qt, which is a higher-order multipole and
hence, in normal circumstances, can be ignored. This situation
suggests that we must go up to an unusually high chiral order before
getting sensible estimates of the isoscalar matrix elements that
govern the spin observables in (\ref{polnp}). However, remarkably,
MEEFT allows us to make a systematic and reasonably reliable
estimation of M1S and E2S completely free of parameters.

We write the transition amplitude as
 \be
 \langle \psi_d,
\gamma(\hatk, \lambda) | {\cal T}|
    \psi_{np}\rangle
 = \chi^\dagger_d\, {\cal M}(\hatk, \lambda)\, \chi_{np}
 \ee
with
 \be {\cal M}(\hatk, \lambda) =&&
 \sqrt{4\pi} \frac{\sqrt{v_n}}{2 \sqrt{\omega} A_s}
 \, \Big[
  i (\hatk \times {\hat \epsilon}_\lambda^*)\cdot (\vs_1-\vs_2)\, \MS
\nonumber \\
  && - i (\hatk \times {\hat \epsilon}_\lambda^*)\cdot
  (\vs_1+\vs_2)\,\frac{\mut}{\sqrt{2}}\nonumber\\
    && + (\vs_1\cdot\hatk \vs_2\cdot {\hat \epsilon}_\lambda^*
   + \vs_2\cdot\hatk \vs_1\cdot {\hat \epsilon}_\lambda^*)
    \frac{\Qt}{\sqrt{2}}
\Big]\label{amp} \ee where $v_n$ is the velocity of the projectile
neutron, $A_s$ is the deuteron normalization factor $A_s\simeq
0.8850\ {\rm fm}^{-1/2}$, and $\chi_d$ ($\chi_{np}$) denotes the
spin wave function of the final deuteron (initial $np$) state. The
emitted photon is characterized by the unit momentum vector $\hatk$,
the energy $\omega$ and the helicity $\lambda$, and its polarization
vector is denoted by ${\hat \epsilon}_\lambda \equiv {\hat
\epsilon}_\lambda({\hat k})$. The amplitudes, $\MS$, $\mut$ and
$\Qt$, represent the isovector M1, isoscalar M1 and isoscalar E2
contributions, respectively.\footnote{These amplitudes are real at
threshold.} These quantities are defined in such a manner that they
all have the dimension of length, and the cross section for the
unpolarized $np$ system takes the form
 \be \sigma_{unpol}=
\abs{\MS}^2 + \abs{\mut}^2 + \abs{\Qt}^2\,. \label{xsection}
 \ee As mentioned above, the isovector M1 amplitude was
calculated~\cite{pmr-np1} very accurately up to ${\cal O} (Q^3)$
relative to the single-particle operator. The result expressed in
terms of $\MS$ is: $\MS{}=5.78\pm 0.03$ fm~\footnote{In the same
notation, the empirical value is $\sqrt{\sigma_{unpol}^{exp}}=
5.781\pm 0.004$ fm.}. So we need to focus on the isoscalar matrix
elements. The isoscalar matrix elements are given by
 \be \mut &\equiv& -
\frac{\sqrt{2}\omega^{\frac32}}{\sqrt{v_n}}
 \langle \psi_d^{J_z=1} | \mu^z | \psi_t^{J_z=1}\rangle,
 \nonumber \\
\Qt &\equiv& \frac{\omega^{\frac52}}{\sqrt{8} \sqrt{v_n}}
  \langle \psi_d^{J_z=1} | Q^{33} | \psi_t^{J_z=1}\rangle
\label{isoscalarME}\ee with \be {\vec \mu}&=& \frac12 \int\! d^3 \vx
\, \vx \times \vJ_\EM(\vx),
\nonumber \\
Q^{ij}&=& \int\! d^3 \vx \, (3 x^i x^j - \delta^{ij} \vx^2)
J_\EM^0(\vx), \label{muQdef}
 \ee where $J_\EM^\mu(\vx)$ is the EM current.
%
Now the measurable quantities are the photon circular polarization
$P_\gamma$ and the anisotropy $\eta_\gamma$ defined in terms of the
angular distribution of photons with helicity $\lambda=\pm 1$
denoted $I_\lambda(\theta)$ where $\theta$ is the angle between
$\hatk$ (direction of photon emission) and a quantization axis of
nucleon polarization. For polarized neutrons with polarization
$\vec{P}_n$ incident on unpolarized protons, $P_\gamma$ is defined
by
 \be P_\gamma \equiv \frac{I_{+1}(0^\circ) - I_{-1}(0^\circ)}
{I_{+1}(0^\circ) + I_{-1}(0^\circ)}\,.\label{pgammadef}
 \ee
With both protons and neutrons polarized (along a common
quantization axis) with polarizations $\vec{P}_n$ and $\vec{P}_p$,
respectively, the anisotropy $\eta_\gamma$ is defined by
 \be
\eta_\gamma &\equiv& \frac{I(90^\circ) - I(0^\circ)} {I(90^\circ) +
I(0^\circ)}\,,
 \ee
where $I(\theta)= I_{+1}(\theta) + I_{-1}(\theta)$ is the angular
distribution of total photon intensity (regardless of their
helicities). These quantities are given in terms of $\vec{P}_{p,n}$
and the ratios of matrix elements
 \be {\cal R}_{\rm M1} \equiv
\frac{\mut}{\MS}\;, \ \ \ {\cal R}_{\rm E2} \equiv
\frac{\Qt}{\MS}\;.\label{ratio}
 \ee
See \cite{pkmr-isonp} for explicit formulas.

The matrix elements M1S and E2S have been computed to ${\cal O}
(Q^4)$ in the chiral counting defined above that involves up to
one-loop graphs. It turns out to the order considered that E2S is
given entirely by the one-body matrix elements, with the two-body
correction estimated to be $\Qt_{\rm 2B} = (0.00\pm 0.01)\
\times10^{-3}\ \fm$ for the whole range of $r_c=(0.01 \sim 0.8)\
\fm$. Combining the one-body and two-body contributions, it is found
to be
 \be \Qt =
(1.40 \pm 0.01) \times 10^{-3}\ \fm.
 \ee
Thus the most interesting quantity is the two-body contribution to
M1S denoted $\mut_{\rm 2B}$.  One naively would think that there
would be too many parameters to the chiral order involved to make a
parameter-free prediction. It turns out however that this is not the
case. There is one unknown parameter that appears at ${\cal O}
(Q^3)$ in the form of a contact counter term denoted $g_4^\prime$ in
\cite{pkmr-isonp} that however can be determined entirely by the
deuteron magnetic moment (which is of course isoscalar). This means
that apart from the cut-off $\Lambda$ or in the coordinate space
$r_c$, there is no unknown parameter in the theory. $\mut_{\rm 2B}$
is of the form
 \be
\mut_{\rm 2B}=a(r_c)+g_4^\prime b(r_c)\label{M1S2B}
 \ee
where both $a(r_c)$ and $b(r_c)$ diverge for $r_c\rightarrow 0$ but
otherwise are completely determined for any $r_c\neq 0$. The second
term comes from the contact (counter) term which is of delta
function in coordinate space. The premise of the consistency of EFT
dictates that the sum of the total MS1=MS1$_{\rm 1B}$+MS1$_{\rm 2B}$
be insensitive to the cut-off $r_c$ within the physically reasonable
range defined above. This is indeed what comes out. For the
wide-ranging value of $r_c=0.01, 0.2, 0.4, 0.6, 0.8$ fm, fitting to
the deuteron magnetic moment requires the corresponding $g_4^\prime$
to be $g_4^\prime=5.06, 2.24, 0.73, 0.31$. Although the $g_4^\prime$
varies strongly in the range considered, the total M1S varies
negligibly: M1S$\times 10^3$ fm$^{-1}$ = $-$(2.849, 2.850, 2.852,
2.856, 2.861). This allows to predict the M1S to a high accuracy:
 \be \mut = (-2.85\pm 0.01)\times 10^{-3}\
 \fm\,,\label{predictMEEFT}
 \ee
where the error bar stands for the $r_c$-dependence.

This accurate prediction has however remained untested
experimentally. We should underline at this point that exactly the
same situation will arise in the solar $pp$ and $hep$ calculation
that will be given below.

It is interesting to compare the surprisingly precise prediction
(\ref{predictMEEFT}) with what one obtains in
$\pislash$EFT\cite{chen-np-supp}. Because of the dominance of the
single-particle matrix element with the next-order corrections
suppressed, one obtains roughly the same $\Qt$ in $\pislash$EFT as
in MEEFT. The situation is quite different for $\mut$ however. Here
due to the ``accidental" suppression of the leading order (one-body)
isoscalar M1 operator, the next-order term, though down formally by
${\cal O}(Q^3)$, is substantially bigger than naively expected. Thus
in the $\pislash$EFT calculation, the leading order correction
cannot be completely pinned down by the deuteron magnetic moment
alone as in the case of MEEFT. There is an undetermined constant
which cannot be taken into account, making the calculation of the
M1S uncertain by $\sim 60$ \% or more. Thus both M1V and M1S are not
really predictable in this approach.

\subsubsection{Deuteron form factors}
An objection may be raised at this point against the above claim of
success on the basis of (in)consistency in the chiral counting. The
highly ``sophisticated" wave function that is used in the
calculation can in principle account for $all$ orders of chiral
expansion in the kinematic regime we are concerned with whereas the
currents are calculated to a finite order N$^n$LO for $n<\infty$.
This means that there is in a strict sense an inconsistency in
chiral counting at order $n+1$. Given that $all$ calculations based
on systematic expansion involve truncation at a certain level of
accuracy, none of what we might consider as ``precise results" can
``rigorously" claim to be perfectly consistent. So the relevant
question is: Is the possible inconsistency serious?

There is no fully satisfactory answer to this question in the case
at hand. Here we would like to show that at least within the
few-body systems we are discussing, we see no evidence for serious
inconsistency in the MEEFT scheme. To illustrate this point, we take
the well-tested case of the electron-deuteron elastic scattering
 \be
e+d\rightarrow e+d\label{edscattering}
 \ee
at low momentum transfers $Q=\sqrt{-q^2} \lsim 0.8$ GeV. This
process has been analyzed both in RigEFT and in MEEFT with the
currents computed to ${\calO}(Q^3)$. At the end of this subsection,
we will comment on how one can understand the well-known ``$Q_d$
problem" in terms of the chiral filter mechanism.

The process (\ref{edscattering}) involves precisely the same EM
current that figured in the polarization observables in $np$ capture
discussed above. It is isoscalar and hence in the terminology
introduced above, multi-body corrections are chiral-filter
unprotected. This means that relative to the leading one-body
current, the corrections are suppressed at least by ${\calO}(Q^2)$.
Beyond that order, short-distance dynamics uncontrollable by chiral
symmetry may play an extremely important role as we saw above and
will encounter again below. As remarked, one expects that it is in
processes which are not protected by chiral filter that the possible
inconsistency in chiral counting of MEEFT, if any, should show up
and have serious consequence. (To repeat, the chiral-filter
protected processes are dominated -- barring accidental suppression
-- by one-pion-exchange contributions and hence the possible error
that could be due to inconsistency is within the error bar -- both
theoretical and experimental.)

Our chief point can be made based on the work by
Phillips~\cite{phillips-formfactor}. In this work, it was found that
the ratios of the form factors provided more useful information on
the working of EFT than the form factors by themselves. Our
discussion will focus on both the comparison between the RigEFT and
MEEFT in their $confronting$ experimental data.

To define notations, we write the differential cross section for
electron-deuteron scattering in the lab frame as
\begin{equation}
\frac{d \sigma}{d \Omega}=\frac{d \sigma}{d \Omega}_{0} \left[A(Q^2)
+ B(Q^2) \tan^2\left(\frac{\theta_e}{2}\right)\right].
\label{eq:dcs}
\end{equation}
Here $\theta_e$ is the electron scattering angle, and $\frac{d
  \sigma}{d \Omega}_{0}$ is the cross
section for electron scattering from a point particle of charge
$|e|$ and mass $M_d$ in one-photon exchange approximation.  The
Coulomb (C), electric quadrupole (Q) and magnetic (M) form factors
figure in $A$ and $B$ as
\begin{eqnarray}
A&=&G_C^2 + \frac{2}{3} \eta G_M^2 + \frac{8}{9} \eta^2  M_d^4
G_Q^2,\\
\label{eq:A} B&=&\frac{4}{3} \eta (1 + \eta) G_M^2. \label{eq:B}
\end{eqnarray}
These form factors are normalized so that at zero momentum transfer,
\begin{eqnarray}
G_C(0)&=&1,\\
G_Q(0)&=&Q_d,\\
G_M(0)&=&\mu_d \frac{M_d}{m_N},
\end{eqnarray}
where $Q_d$ is the quadrupole moment and $\mu_d$ the magnetic moment
of the deuteron. The ratios in question
\begin{equation}
\frac{G_C}{G_E^{(s)}} \quad \mbox{and} \quad \frac{G_Q}{G_E^{(s)}}
\quad \mbox{and} \quad \frac{G_M}{G_M^{(s)}}, \label{eq:ratios}
\end{equation}
with $G_E^{(s)}$ and $G_M^{(s)}$ the isoscalar single-nucleon
electric and magnetic form factors, are argued to be better behaved
than the chiral expansion for $G_C$, $G_Q$, and $G_M$ themselves.

We consider in particular the results of the calculation performed
to ${\calO}(Q^3)$ for the potential and for the current in
one-photon-exchange approximation~\footnote{One can count this as we
have done above for the currents as relative to the leading order
charge operator which is of ${\calO}(Q^0)$.}. At higher orders, the
$g_4^\prime$-type corrections enter crucially in $G_M$ which we will
not consider.

As given, we mean by RigEFT the calculation of the form factors with
the wave functions obtained with a potential computed up to
${\calO}(Q^3)$ and the currents computed to the same order. This
represents a fully consistent EFT. By MEEFT, we mean calculating the
matrix elements with the current computed to order ${\calO}(Q^3)$
and the wave functions obtained with SNPA potentials -- meaning the
``high-quality realistic potentials." The SNPA potentials used by
Phillips are the Nijm93\cite{nijm93} and CD-Bonn\cite{cd-bonn}
potentials. These potentials have the common feature of having
correct long-range part (i.e., one pion exchange) with differing
short-range part but fit to NN scattering to momenta $\sim 2{\rm
fm}^{-1}$. This feature is shared by other high-quality potentials
such as the Argonne $v_{18}$ potential\cite{av18} used in the solar
neutrino processes described below. We take the Nijm93 and CD-Bonn
potentials as representative ``accurate" potentials.

%
%

It is clear in comparing various calculations described in detail in
\cite{phillips-formfactor} and the experimental data that due to
large error bars of the experimental form factors, one cannot
discriminate the optimal RigEFT results and the MEEFT results. Nor
is it possible to gauge whether there is any serious inconsistency
in the EFT counting rule which is, strictly speaking, inevitable in
MEEFT even if it turns out to be ignorable. Indeed the uncertainty
in MEEFT which may be manifested in the difference between two
``reliable" potentials employed in \cite{phillips-formfactor}, here
represented by the Nijm93 and CD-Bonn potentials, is shown to be
considerably less than the uncertainty manifested in different
treatments of higher-order terms in the RigEFT approach. In fact,
within the present precision of the experimental data, it is
perfectly reasonable to conclude that the MEEFT fares equally well
as, if not much better than, the RigEFT result in explaining the
experimental data. We repeat that this is a case of the
chiral-filter unprotected processes which is the least favorable to
MEEFT due to possible counting inconsistency.

One lesson we can draw from the consideration of the deuteron EM
form factors is that it illustrates the subtle nature of the chiral
filter mechanism, manifested as the ``two sides of the same coin."
Consider for this the E2 response of deuterium as we discussed above
for both the polarized $np$ capture process and the deuteron EM form
factors. As noted above, the E2 matrix element (denoted E2S) for the
$np$ capture is calculable in both $\pislash$EFT and MEEFT up to the
chiral order that involves no free parameters (i.e., ${\cal O}(Q^3)$
in MEEFT). This is easy to see with the pions included. One notes
that the $1/M$ corrections which come at ${\cal O}(Q^3)$ are
essentially governed by chiral symmetry (i.e., pions) and Poincar\'e
invariance and hence more or less
model-independent~\cite{phillips-formfactor}. Furthermore the
next-order terms, i.e., of ${\cal O}(Q^4)$, are also calculable
parameter-free as shown by Park et al~\cite{pkmr-isonp}. However the
resulting prediction for the deuteron quadrupole moment $Q_d$
undershoots the experiment by 2-3\%. This is a ``huge" discrepancy
considering the accuracy achieved in the M1 matrix elements. As
suggested by Phillips~\cite{phillips-formfactor}, the possible
solution to this ``$Q_d$ problem" is in zero-range ${\cal O}(Q^5)$
terms -- with, however, undetermined coefficients -- that represent
short-distance physics in a way analogous to the $g_4^\prime$ term
in the isoscalar M1 transitions discussed above and the $\dR$ term
in the isovector Gamow-Teller transitions discussed below. What is
common in all the cases considered here is that the single-particle
matrix elements are unnaturally suppressed, making correction terms
particularly important, sometimes even dominant. Now, when
chiral-filter protected, the leading corrections dictated by chiral
symmetry dominate and hence account for most, if not all, of the
required matrix elements. However when chiral-filter unprotected as
in the $Q_d$ case in question, even though one may be able to
calculate free of parameters one or two next-order corrections which
are intrinsically suppressed to start with, they do not saturate the
corrections. One has to go to an arbitrarily high order generated by
short-distance interactions before the corrections can, if at all,
be put under control.
\subsubsection{Prediction of certain solar neutrino processes}
Now we are ready to make a precise statement on the key prediction
of the MEEFT approach that has not yet been matched by the RigEFT
strategy. They are the solar $pp$ and $hep$ processes. We note here
that the calculation of the $hep$ process has not yet been achieved
by the RigEFT method.~\footnote{The challenge made in 2000 at a
meeting in Taipei (not put in the written version~\cite{taiwan}) to
the aficionados of the $\pislash$EFT to come up with a comparable
(parameter-free) prediction for the $hep$ process has remained, as
far as the author is aware, yet unmet.}

To treat the axial-current entering in these processes, we need to
fix the unknown constants in the axial two-body operator
(\ref{vAnuFT}) $\vA_{12}(\vr)$. Now in $\vA_{12}(\vr)$,
$\hat{c}_{3,4}$ are fixed from $\pi-N$ scattering, so that leaves
only one parameter to be fixed, i.e., $\dR$, multiplying the delta
function term. Actually there are two unknown terms $\hat{d}_{1,2}$
in $\vA_{12}(\vr)$ originating from a term of the form in the
current
 \be
- \frac{g_A}{m_N f_\pi^2} \left[
 2 \hat d_1 (\tau_1^- \vs_1 + \tau_2^- \vs_2)
 + \hat d_2 \tau^a_\times \vs_\times
 \right] \>\>
 \ee
but it turns out from Fermi statistics that only one combination
 \be
 \dR\equiv \hat d_1 +2 \hat d_2 +
\frac13 \hat c_3
 + \frac23 \hat c_4 + \frac16 \>\>
\label{dr}
 \ee
figures in all the relevant processes that we are concerned with.
This parameter $\dR$ is analogous to $g_4^\prime$ that figures in
the isoscalar $np$ capture treated above.  The argument why this is
the only relevant combination is given in the paper \cite{hep} which
should be consulted for details. It follows from symmetry
considerations.

Once $\dR$ is fixed by fitting a well-measured quantity in a
specific system, i.e., the triton beta decay in the present case,
then it is fixed {\it once and for all independently of the mass
number} involved. The corresponding matrix element is determined for
$any$ system given the wave functions. The reason for the
universality of $\dR$ (and also $g_4^\prime$) is that the
corresponding operator is short-ranged and hence should be
independent of the density of the system as long as the Fermi
momentum is much less than the cutoff. If there were uncertainty due
to regularization which would be the case if there is strong
mismatch in the chiral counting between the wave functions which are
obtained by empirical fits and the current operators which are
computed to a certain order in the chiral counting, then physical
quantities would exhibit dependence on the cutoff, that is, on the
regulator. Thus an a posteriori justification of the procedure would
be the cutoff independence. This is not a rigorous justification but
it is the best one can do in any effective theories which are by
definition an approximation.

We illustrate how this strategy works {\it in making parameter-free
predictions} with the second (and third) and fifth processes of
(\ref{meeft}) relevant to solar neutrino problems.
\subsubsection{The $pp$ process}
Here we focus on the $pp$ process. The same strategy applies to the
processes (\ref{CC}) and (\ref{NC}) involving
neutrinos~\cite{andoetal-neutrino}. It is convenient to decompose
the matrix element of the GT operator into one-body and two-body
parts
 \be \calM= \calM_{\rm
1B} + \calM_{\rm 2B} \>\>. \ee Since the one-body term is
independent of the cutoff and very well known in SNPA, we discuss
only the two-body term.


The properly regularized two-body GT matrix elements for the $pp$
process read
 \be \calM_{\rm 2B} &=&
 \frac{2}{m_N f_\pi^2}
 \int_0^\infty\! dr\,  \left\{ \frac{}{} \right.
\nonumber \\
&&
 \frac{m_\pi^2}{3}
\left(\hat c_3 + 2 \hat c_4 + \frac12\right)
  y_{0\Lambda}^\pi(r)\, u_d(r)\,  u_{pp}(r)
\nonumber\\
&-& \sqrt{2} \frac{m_\pi^2}{3}
 \left(\hat c_3 - \hat c_4 - \frac14\right)
 y_{2\Lambda}^\pi(r)\, w_d(r) \, u_{pp}(r)
\nonumber \\
&+&
 \frac{y_{1\Lambda}^\pi(r)}{12 r} \Bigg[
 \left[ u_d(r)-\sqrt2 w_d(r)\right]
 u_{pp}'(r) \nonumber \\
&-& \left[ u_d'(r)-\sqrt2 w_d'(r)\right] u_{pp}(r) +
\frac{3\sqrt2}{r} w_d(r) u_{pp}(r)
\Bigg] \nonumber \\
&-&\left. \dR \delta_\Lambda^{(3)}(r)\, u_d(r) u_{pp}(r)
\frac{}{}\right\}, \label{calM2Bdelta} \ee where $u_d(r)$ and
$w_d(r)$ are the S- and D-wave components of the deuteron wave
function, and $u_{pp}(r)$ is the $^1$S$_0$ $pp$ scattering wave (at
zero relative energy). The results are given for the three
representative values of $\Lambda$ in Table~\ref{tb:tb1}; for
convenience, the values of $\dR$ given in Eq.(\ref{dR}) are also
listed. The table indicates that, although the value of $\dR$ is
sensitive to $\Lambda$, $\calM_{\rm 2B}$ is amazingly stable against
the variation of $\Lambda$ within the stated range. In view of this
high stability, we believe that we are on the conservative side in
adopting the estimate $\calM_{\rm 2B}= (0.039 \sim 0.044)\ \fm$.
Since the leading single-particle term is independent of $\Lambda$,
the total amplitude $\calM = \calM_{\rm 1B}+\calM_{\rm 2B}$ is
$\Lambda$-independent to the same degree as $\calM_{\rm 2B}$. The
$\Lambda$-independence of the physical quantity $\calM$ is a crucial
feature of the result in our present study. The relative strength of
the two-body contribution as compared with the one-body contribution
is \be \delta_{\rm 2B} \equiv \frac{\calM_{\rm 2B}}{\calM_{\rm 1B}}
= (0.86 \pm 0.05)\ \%. \label{delta2B-new}
 \ee
Despite that this process is not protected by the chiral filter, we
have achieved an accuracy unprecedented in nuclear physics. This
aspect will be exploited for the case of the $hep$ process below.
\begin{table}[hb]

{\begin{tabular}{|c|c|l|}\hline
$\Lambda$ (MeV) & $\dR$ & $\calM_{\rm 2B}$ (fm) \\
\hline 500 & $1.00 \pm 0.07$ &
$0.076 - 0.035\ \dR \simeq 0.041 \pm 0.002 $\\
\hline 600 & $1.78 \pm 0.08$ & $0.097 - 0.031\ \dR \simeq 0.042 \pm
0.002$
\\ \hline
800 & $3.90 \pm 0.10$ & $0.129 - 0.022\ \dR \simeq 0.042 \pm 0.002$
\\  \hline
\end{tabular}}\caption{\protect The strength $\dR$ of the contact term and the
two-body GT matrix element, $\calM_{\rm 2B}$, for the $pp$ process
calculated for representative values of $\Lambda$.}\label{tb:tb1}
\end{table}

To be complete and useful to solar neutrino studies, we write down
the threshold $S$ factor, $S_{pp}(0)$. Adopting the accurately
determined value $G_V=(1.14939 \pm 0.00065)\times 10^{-5}\
\mbox{GeV}^{-2}$, we obtain \be S_{pp}(0) &=& 3.94\times
\left(\frac{1+\delta_{2B}}{1.01}\right)^2
\left(\frac{g_A}{1.2670}\right)^2
\left(\frac{\Lambda_{pp}^2}{6.91}\right)^2
\nonumber \\
&=&
 3.94\times
  (1 \pm 0.0015 \pm 0.0010 \pm \epsilon)
  \label{S-factor}
\ee in units of $10^{-25}\ \mbox{MeV-b}$. Here the first error is
due to uncertainties in the input parameters in the one-body part,
while the second error represents the uncertainties in the two-body
part; $\epsilon (\approx 0.001)$ denotes possible uncertainties due
to higher chiral order contributions. To make a formally rigorous
assessment of $\epsilon$, we must evaluate loop corrections and
higher-order counter terms. Although an ${\cal O} (Q^4)$ calculation
would not involve any new unknown parameters, it is a non-trivial
task. Furthermore, loop corrections necessitate a more elaborate
regularization scheme since the naive cutoff regularization used
here violates chiral symmetry at loop orders. (This difficulty,
however, is not insurmountable.) These formal problems
notwithstanding, it is possible to give a reasonable justification
of the small correction $\epsilon(\approx 0.001)$ assigned to the
S-factor.

It is somewhat surprising that the short-range physics is so well
controlled in MEEFT at the order considered. In the conventional
treatment of MEC, one derives the coordinate space representation of
a MEC operator by applying ordinary Fourier transformation (with no
restriction on the range of the momentum variable) to the amplitude
obtained in momentum space; this corresponds to setting
$\Lambda=\infty$ in Eq.(\ref{regulator}). Short-range correlation
has to be implemented in an ad hoc manner to account for the
short-distance ignorance. In MEEFT, the inclusion of the $\dR$ term,
with its strength renormalized as described here, properly takes
into account the short-range physics inherited from the
integrated-out degrees of freedom above the cutoff, thereby
drastically reducing the undesirable (or unphysical) sensitivity to
short-distance physics. It is undoubtedly correct to say that this
procedure is not rigorously justified on the basis of strict chiral
counting but it should be stressed that what we might call ``ansatz"
as it stands is a giant leap from the SNPA prescription for
``short-range correlation."
\subsubsection{The $hep$ process}
This process is a lot more complicated than the $pp$ process
involving up to four nucleons, both in bound and scattering states.
To make contact with the literature and also to avoid crowding with
unilluminating formulas, we use the notation of the classic SNPA
paper (called MSVKRB here)~\cite{MSVKRB}, focusing on the essential
part of our MEEFT strategy. The GT-amplitudes will be given in terms
of the reduced matrix elements, $\overline{L}_1(q;A)$ and
$\overline{E}_1(q;A)$.~\footnote{Being more specific about these
matrix elements is not required for our discussion here. What
matters is the behavior of the typical matrix element involved vs.
cutoff.} Since these matrix elements are related to each other as
$\overline{E}_1(q;A) \simeq \sqrt{2}\, \overline{L}_1(q;A)$, with
the exact equality holding at $q$=0, we consider here only one of
them, $\overline{L}_1(q;A)$. For the three exemplary values of
$\Lambda$, Table~\ref{TabL1A} gives the corresponding values of
$\overline{L}_1(q;A)$ at $q\equiv|\vq|$=19.2 MeV and zero c.m.
energy; for convenience, the values of $\dR$ in Eq.(\ref{dR}) are
also listed. We see from the table that the variation of the
two-body GT amplitude (row labelled ``2B-total'') is only $\sim$10
\% for the range of $\Lambda$ under study.  Note that the
$\Lambda$-dependence in the total GT amplitude is made more
pronounced by the drastic cancellation between the one-body and
two-body terms, but this amplified $\Lambda$-dependence still lies
within acceptable levels.

\begin{table}[htb]
{\begin{tabular}{c|rrr}\hline $\Lambda$ (MeV) & 500 & 600 & 800 \\
\hline $\dR$      & $1.00 \pm 0.07$ & $1.78\pm 0.08$ & $3.90\pm
0.10$
\\ \hline
$\overline{L}_1(q;A)$ & $-0.032$ & $-0.029$ & $-0.022$ \\ \hline 1B
& $-0.081$ & $-0.081$ & $-0.081$ \\ \hline
2B (without $\dR$)    & $0.093$  & $0.122$  & $0.166$  \\
2B ($\propto \dR$)    & $-0.044$ & $-0.070$ & $-0.107$ \\ \hline
2B-total              & $0.049$  & $0.052$  & $0.059$  \\ \hline
\end{tabular}}\caption{\protect Values of $\dR$ and $\overline{L}_1(q;A)$ (in
fm$^{3/2}$) for the $hep$ process calculated as functions of the
cutoff $\Lambda$. The individual contributions from the one-body
(1B) and two-body (2B) operators are also listed.}\label{TabL1A}
\end{table}

%

Summarizing the results obtained, we arrive at a prediction for the
$hep$ $S$-factor:
 \be
S_{hep}(0)=(8.6 \pm 1.3 ) \times \Sunit\,,\label{prediction} \ee
where the ``error" spans the range of the $\Lambda$-dependence for
$\Lambda$=500--800 MeV.
\subsubsection{Confront nature}
There is no laboratory information on the $hep$ process and the only
information we have at present is the analysis of the
Super-Kamiokande data \cite{SK2001} which gives an upper limit of
the solar $hep$ neutrino flux, $\Phi(hep)^{\rm SK} < 40 \times 10^3$
cm$^{-2}$s$^{-1}$. The standard solar model \cite{BP2000} using the
$hep$ $S$-factor of MSVKRB~\cite{MSVKRB} predicts $\Phi(hep)^{\rm
SSM} = 9.4 \times 10^3$ cm$^{-2}$s$^{-1}$. The use of the central
value of our estimate, Eq.(\ref{prediction}), of the $hep$
$S$-factor would slightly lower $\Phi(hep)^{\rm SSM}$ but with the
upper limit compatible with $\Phi(hep)^{\rm SSM}$ in
Ref.~\cite{BP2000}. A significantly improved estimate of
$S_{hep}(0)$ in Eq.(\ref{prediction}) is expected to be useful for
further discussion of the solar $hep$ problem.

One can reduce the uncertainty in Eq.(\ref{prediction}). To do so,
one would need to reduce the $\Lambda$-dependence in the two-body GT
term. By the EFT strategy, the cutoff dependence should diminish as
higher order terms are included. In fact, the somewhat rapid
variation seen in $\dR$ and in the ${}^3S_1$ contribution to
$S_{hep}(0)$ as $\Lambda$ approaches 800 MeV may be an indication
that there is need for higher chiral order terms or alternatively
the explicit presence of the vector-mesons ($\rho$ and $\omega$)
with mass $m_V \lsim \Lambda$. We expect that the higher order
correction would make the result for $\Lambda=800$ MeV closer to
those for $\Lambda=500, 600$ MeV. This possibility is taken into
account in the error estimate given in Eq.(\ref{prediction}).

The $hen$ process
 \be
n+^3{\rm He}\rightarrow ^4{\rm He} + \gamma.
 \ee
contains both features of chiral-filter protected and unprotected
operators and could provide a beautiful testing ground for the main
ideas put forward in this paper. The process is mediated by the EM
current which while dominated by isovector M1 operators, gets
non-negligible contributions from isoscalar currents. This is
because as in the $hep$ process, the leading one-body operators are
strongly suppressed by the pseudo-orthogonality of initial and final
wave functions, making multi-body corrections become much more
pronounced. In fact, one can make a rough estimate that the two-body
corrections will dominate while three-body and four-body corrections
are expected to be negligible and hence can be safely ignored within
the accuracy we desire. This process presents a particularly
significant case where two-body corrections are mandatory and are to
be calculated with high accuracy. Now as in the thermal $np$
capture, the two-body isovector M1 operator is dominated by the
chiral-filter protected one-pion exchange graph. However since the
single-particle term is suppressed, one would have to compute next
order corrections to the two-body operator which come at N$^2$LO
relative to the two-body term. At that order, one has a
delta-function counter term analogous to the isoscalar $g_4^\prime$
encountered in the polarization observables in the $np$ capture
discussed above.

Furthermore unlike in the $np$ total capture cross section,
isoscalar contributions cannot be ignored for the $hen$ cross
section. Here because of the same pseudo-orthogonality of the wave
functions, the one-body isoscalar current is doubly suppressed, one
by the small isoscalar nucleon magnetic moment and the other by the
wave function mismatch. This means that a higher-order two-body
isoscalar counter term analogous to the $g_4^\prime$ term needs to
be accounted for. Thus for a high-accuracy calculation, two contact
term coefficients, one for the isovector term and the other for the
isoscalar term, must be determined from given experiments. Most
fortunately, we have at our disposal two accurate experimental data,
namely, the magnetic moments of $^3$He and $^3$H, that remove
completely the dependence on free parameters. Thus it is possible to
make a genuine prediction that can be confronted by the presently
available experimental results.~\footnote{The present experimental
values have $\sim 10$\% error bars. We expect that the theoretical
error bar will be considerably smaller than this and hence the
confrontation can be made a crucial test of the approach if the
experimental error bars can be appreciably reduced.} As observed in
all cases discussed above, there would be a wide variation on the
$hen$ cross section depending on the cutoff values when the counter
terms are not included but the dominance of the chiral-filter
protected two-body correction in the isovector channel would make
the calculated result much less sensitive to the cutoff than in the
$hep$ case when the contact terms are incorporated.
\subsubsection{Further implications of the $\dR$ term} 
\begin{figure}[ht]
\center{\includegraphics[width = 14cm]{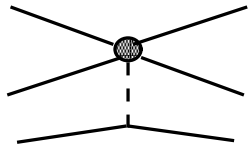}} \vskip -5.5cm
\caption{\label{3-body} The dominant short-ranged 3-body force
mediated by the $\dR$ term (indicated by the blob) exchanging a pion
with the third nucleon. Here the solid lines stand for nucleon
lines, and the broken line for the pion.}
\end{figure}

The
short-range two-body axial current proportional to $\dR$ in
(\ref{vAnuFT}) can intervene in an important way in  different
processes. It is a four-Fermi axial current, so it couples to an
axial field, internal or external. For instance, it can couple to
the pion via PCAC and this led G{\aa}rdestig and
Phillips~\cite{gardestig-phillips} to suggest that one can get the
three-body force that governs short-distance interaction as given by
Fig. \ref{3-body}. This will be able to pin down the three-body
force with a fairly good accuracy. An example that involves external
fields that we will discuss here is the process of the type
 \be
\gamma NN \leftrightarrow \pi NN
 \ee
where the pion involved is soft. This process -- which is a two-body
analog to the Kroll-Ruderman term in photo-pion production on a
nucleon -- can be used to extract reliably the neutron-neutron
scattering length and here the $\dR$ term plays the same role as in
the solar neutrino processes discussed above to assure a model
independent MEEFT. It is an extremely interesting case that
illustrates the power of MEEFT.

It has been shown in \cite{gardestig-phillips} that the uncertainty
in the neutron-neutron scattering length can be reduced to $\lsim
0.05$ fm. Let us discuss how this comes about.

Data from radiative pion capture experiments
 \be
\pi^- + d\rightarrow n+n+\gamma\label{pioncapture}
 \ee
dominate the accepted value
 \be
a_{nn}=-18.59\pm 0.40\ \ {\rm fm}.
 \ee
Now it turns out that $\pm 0.3$ fm out of $\pm 0.4$ fm in the error
bar arises from the short-distance uncertainty and hence in order to
reduce the error substantially below the $\pm 0.3$ fm, it is
necessary to control the short-distance component of the force. It
is here that the $\dR$ term accurately determined above can come in.
The basic idea goes as follows~\cite{gardestig-phillips}. For soft
pion, the process (\ref{pioncapture}) goes through the same current
as the solar $pp$ process (apart from the Coulomb interaction) via
PCAC. The relevant Lagrangian is of the form
 \be
\delta {\calL}=-\frac{2\hat{d}_1}{m_N f_\pi^2}N^\dagger S\cdot u N
N^\dagger N.
 \ee
The coefficient $\hat{d}_1$ here is, apart from a known constant,
essentially the soft-pion limit of the coefficient $\dR$, so the
idea is to determine the coefficient $\hat{d}_1$ for a given
short-distance scale parameter which in the case of the $pp$ process
was the cutoff $\Lambda$. Then using the same prescription for
short-distance treatment, one computes the capture process
(\ref{pioncapture}) using the $\hat{d}_1$ in a suitable range of the
short-distance cutoff parameter. The crucial idea is that an
approximate renormalization group invariance (in the Wilsonian
sense) is achieved if within the reasonable range of cutoff
parametes, the physical observable is insensitive to the change of
the cutoffs. The procedure taken in \cite{gardestig-phillips} to
implement this strategy was as follows. For both the deuteron and
scattering wave functions, they are calculated from $r=\infty$ to a
matching radius $R$ -- which plays the role of delineating short
from long distance -- using the one-pion exchange potential and for
$r<R$, a spherical well potential is assumed, the wave function of
which is matched to the $r>R$ wave function. This procedure is
applied to both the $pp$ process and the pion capture
(\ref{pioncapture}), with the Coulomb effect suitably taken into
account into the former. The coefficient $\hat{d}_1$ is an important
part of the $\dR$ that governs the $pp$ process as given by the
result (\ref{S-factor}). The result comes out to be
 \be
\hat{d}_1 = && -1.27 \ {\rm for}\ R=1.4\ {\rm fm}, \  0.48 \ {\rm
for}\ R=2.2\ {\rm fm},\nonumber\\
&&  4.29 \ {\rm for}\ R=3.0\ {\rm fm}.
 \ee
\begin{figure}[hbt]
\center{\includegraphics[width = 8cm]{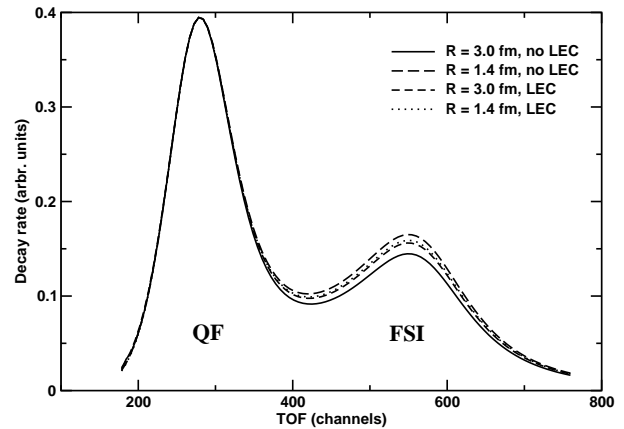}}
\caption{The $\pi^-d\to nn\gamma$ neutron time-of-flight
distribution calculated by G{\aa}rdestig and
Phillips~\cite{gardestig-phillips}, for two widely-separated $R$
values, without and with the low-energy constant (LEC) $\hat{d}_1$
contribution.
The labels QF (quasi-free) and FSI indicate where the corresponding
kinematics are dominant. It is in FSI that the $\hat{d}_1$
contribution makes a prominent effect.} \label{fig:pidR}
\end{figure}
As in the solar neutrino case, the $\hat{d}_1$ varies widely for a
range of the matching radius but the result for the radiative
capture process (\ref{pioncapture}) comes out to be highly
insensitive to $R$. In Fig. \ref{fig:pidR}~\footnote{We would like
to thank the authors of \cite{gardestig-phillips} for permitting us
to reproduce this figure.} is shown the capture rate for $R$ ranging
widely from 1.4 fm to 3.0 fm. This illustrates how well the
short-distance physics is captured by the simple procedure as in the
$pp$ case: while the results differ appreciably for the two widely
different values of $R$ without the $\hat{d}_1$ term in the ``FSI"
region, they are very close -- nearly indistinguishable -- when the
$\hat{d}_1$ term is taken into account. A careful analysis of
\cite{gardestig-phillips} suggests that this procedure will be able
to reduce the short-distance uncertainty in the determination of the
$nn$ scattering length by a factor of 3 relative to the value
available up to date, to, say, $\lsim$ 0.05 fm.
\section{The EFT ``completion" of SNPA} One can now pin-point how EFT
comes in to $complete$ the SNPA. This can be illustrated by the
results of the $hep$ process. In SNPA approaches,  two-body
corrections have been made in a procedure suggested a long time ago
in \cite{CR71}. As discussed in \cite{hep}, there are two serious
problems with this procedure. The first is that without the guidance
of the EFT counting rule, terms of various chiral order are mixed in
at the level of computing leading-order exchange current terms. This
raises the question of consistency. Since the corrections are
chiral-filter unprotected, this inconsistency in chiral counting
could generate serious errors. The second is more serious, having to
do with the shorter-range interaction of the $\dR$ type. In the
procedure of the SNPA, short-range interactions with the range
shorter than the hard-core radius $r_c$ are killed by short-range
correlations. This simply means that roughly speaking, terms of the
$\dR$ type are missing. Without the MEEFT strategy, what enters as
corrections would be largely arbitrary. One can see from Table
\ref{TabL1A} that without the balance from the $\dR$ term, the
corrections are strongly dependent on the cut-off $r_c$ and hence
totally arbitrary, with the result differing by several factors for
different cut-offs. This arbitrariness is neatly circumvented by the
MEEFT procedure at the level where the calculation can be done
without unknown parameters.

The $\dR$ effect can of course be incorporated in a RigEFT with the
pion field retained. Given that what matters is the proper account
of the long-wavelength physics governed by chiral symmetry and of
the short-distance physics subsumed in the ${\cal O}(Q^3)$ terms,
the same set of systems could be treated in a pionful RigEFT and a
prediction (parameter-free) could be made. This exercise has not
been done yet and it would be interesting to see to what extent the
``high accuracy" in SNPA wave functions improves or worsens the
prediction and to what extent there is inconsistency -- if any -- in
chiral counting in the MEEFT procedure.
\section{EFT for heavy nuclei and nuclear matter}
So far we have discussed how to exploit EFT in few-nucleon systems.
Can one extend the same strategy to heavier nuclei on the one hand
and to nuclear matter on the other? Ultimately we would like to go
to densities higher than that of nuclear matter so as to enter the
regime where phase transitions to other forms of states~\cite{CND2}
(such as kaon condensation, quark matter etc.) are supposed to take
place.

In RigEFT, one might try to approach nuclear matter by doing
high-order $\chi$PT starting with a chiral Lagrangian appropriate
for $n$-body systems but defined at zero density (i.e., ``free
space") by systematically including diagrams involving $n$ nucleons.
A possible scenario -- similar in spirit to the double decimation
approach mentioned below and in \cite{BR:DD} -- was suggested along
this line by Lynn~\cite{lynn} which consisted of first constructing
a non-topological soliton, called ``chiral liquid" and then do
fluctuations around the soliton. This program has not yet been
successfully effectuated and it is not even clear that it is doable
in practice. Now in a much more practical way, one might extend the
MEEFT method we described above for few-nucleon systems to
many-nucleon systems with the ``realistic" two-body potentials used
efficiently in the few-body problems supplemented with three-body
and perhaps more-body potentials. The standard nuclear physics
approach belongs to this class of approaches but, so far, without
the proper incorporation of chiral symmetry. The question then is:
How should one go about formulating an MEEFT that is applicable to
heavy nuclei and nuclear matter?

This question is addressed elsewhere~\cite{BR:DD,BHLR06}
after introducing the notion of hidden local symmetry in which
vector mesons enter importantly as relevant degrees of freedom. Here
we mention merely that when one approaches the nuclear matter
saturation density with the intention of going beyond the nuclear
matter density, it is much more astute to approach the problem via
multiple decimations in the renormalization group sense rather than
going in one step as one does in the standard $\chi$PT approach.

Just to illustrate the basic idea, consider the density regime near
the nuclear saturation density, $n \sim 0.16$ fm$^{-3}$. (Things can
become considerably subtler at higher densities.)
Now what characterizes nuclear matter is the presence of the Fermi
surface. It is understood that many-body interactions in the
presence of Fermi surfaces generically lead -- with certain
exceptions that we are not concerned with here -- to an RG fixed
point, known as ``Fermi liquid fixed point"\cite{shankar}.
This means that starting from a chiral Lagrangian defined at zero
density, it makes a good sense to do the first decimation from the
chiral scale $\Lambda_\chi \sim 4\pi f_\pi\sim 1$ GeV to the scale
at which the effective interactions between nucleons in medium are
defined, say, $\Lambda_{eff}\sim 2-3 m_\pi$ and then make the second
decimation to the Fermi liquid fixed point by going from
$\Lambda_{eff}$ to the Fermi surface. In both of these decimations,
Brown-Rho scaling (recently reviewed in \cite{BHLR06}) is found to
play an essential role. It will turn out that this procedure is
consistent with the vector manifestation phenomenon of hidden local
symmetry theory described in \cite{HY:PR,BR:DD,BHLR06}.
The bridge to dense matter where chiral phase transition can take
place is then made through a mapping to hidden local symmetry with
the vector manifestation.
\vskip 0.5cm \centerline{\bf Acknowledgments}

This paper, a part of the book I am writing with Chang-Hwan Lee at
Pusan National University, reviews the work done over many years in
collaboration with Gerry Brown, Kuniharu Kubodera, Dong-Pil Min and
Tae-Sun Park to whom I am deeply indebted. The hospitality of PNU
where this note was written should be acknowledged. I am very
grateful for helpful comments on this note from Kuniharu Kubodera,
Daniel Phillips and Steven Weinberg. Needless to say, none of my
collaborators should be held responsible for possible errors or
misinterpretations that I might be making here. In addition, some of
the ideas developed here may not be fully shared by them.


\end{document}